\documentclass[fleqn,usenatbib]{mnras}

\usepackage{newtxtext,newtxmath}
\usepackage{mdframed}

\usepackage[T1]{fontenc}
\usepackage{ae,aecompl}

\usepackage{graphicx}	% Including figure files
\usepackage{amsmath}	% Advanced maths commands
\usepackage{todonotes}
\usepackage[displaymath, mathlines,running, switch, columnwise]{lineno}

\usepackage[capitalize]{cleveref}
\usepackage{url}
\Crefname{equation}{Equation}{Equations}
\crefname{equation}{Eq.}{Eqs.}
\Crefname{figure}{Figure}{Figures}
\crefname{figure}{Fig.}{Figs.}
\Crefname{table}{Table}{Tables}
\crefname{table}{Tab.}{Tabs.}
\Crefname{section}{Section}{Sections}
\crefname{section}{Sec.}{Secs.}

\def\LCDM{$\Lambda$CDM}
\def\Planck{\textit{Planck} }
\def\P{\mathcal{P}}
\def\L{\mathcal{L}}
\def\Z{\mathcal{Z}}
\def\rmd{\mathrm{d}}
\def\mnest{\texttt{MultiNest}}
\def\pchord{\texttt {PolyChord}}
\def\cosmosis{\texttt {CosmoSIS}}
\def\nlive{${\tt n_{live}}$}
\def\nrepeats{${\tt n_{repeats}}$}
\newcommand{\tol}{${\tt tolerance}$}
\newcommand{\eff}{${\tt efficiency}$}

\title[Robust sampling for the Dark Energy Survey]{Robust sampling for weak lensing and clustering analyses with the Dark Energy Survey}

\author[DES Collaboration]{
\parbox{\textwidth}{
\Large
P.~Lemos,$^{1,2}$\thanks{E-mail: p.lemos@sussex.ac.uk}
N.~Weaverdyck,$^{3,4}$\thanks{E-mail: nweaverdyck@lbl.gov} 
R.~P.~Rollins,$^{5}$
J.~Muir,$^{6}$
A.~Fert\'e,$^{7}$
A.~R.~Liddle,$^{8}$
A.~Campos,$^{9}$
D.~Huterer,$^{3}$
M.~Raveri,$^{10}$
J.~Zuntz,$^{11}$
E.~Di Valentino,$^{5}$
X.~Fang,$^{12,13}$
W.~G.~Hartley,$^{14}$
M.~Aguena,$^{15}$
S.~Allam,$^{16}$
J.~Annis,$^{16}$
E.~Bertin,$^{17,18}$
S.~Bocquet,$^{19}$
D.~Brooks,$^{1}$
D.~L.~Burke,$^{20,21}$
A.~Carnero~Rosell,$^{22,15,23}$
M.~Carrasco~Kind,$^{24,25}$
J.~Carretero,$^{26}$
F.~J.~Castander,$^{27,28}$
A.~Choi,$^{29}$
M.~Costanzi,$^{30,31,32}$
M.~Crocce,$^{27,28}$
L.~N.~da Costa,$^{15,33}$
M.~E.~S.~Pereira,$^{34}$
J.~P.~Dietrich,$^{19}$
S.~Everett,$^{35}$
I.~Ferrero,$^{36}$
J.~Frieman,$^{16,37}$
J.~Garc\'ia-Bellido,$^{38}$
M.~Gatti,$^{10}$
E.~Gaztanaga,$^{27,28}$
D.~W.~Gerdes,$^{39,3}$
D.~Gruen,$^{40,19}$
R.~A.~Gruendl,$^{24,25}$
J.~Gschwend,$^{15,33}$
G.~Gutierrez,$^{16}$
S.~R.~Hinton,$^{41}$
D.~L.~Hollowood,$^{35}$
K.~Honscheid,$^{42,43}$
D.~J.~James,$^{44}$
K.~Kuehn,$^{45,46}$
N.~Kuropatkin,$^{16}$
M.~Lima,$^{47,15}$
M.~March,$^{10}$
P.~Melchior,$^{48}$
F.~Menanteau,$^{24,25}$
R.~Miquel,$^{49,26}$
R.~Morgan,$^{50}$
A.~Palmese,$^{12}$
F.~Paz-Chinch\'{o}n,$^{24,51}$
A.~Pieres,$^{15,33}$
A.~A.~Plazas~Malag\'on,$^{48}$
A.~Porredon,$^{42,43}$
E.~Sanchez,$^{52}$
V.~Scarpine,$^{16}$
M.~Schubnell,$^{3}$
S.~Serrano,$^{27,28}$
I.~Sevilla-Noarbe,$^{52}$
M.~Smith,$^{53}$
E.~Suchyta,$^{54}$
M.~E.~C.~Swanson,$^{55}$
G.~Tarle,$^{3}$
D.~Thomas,$^{56}$
C.~To,$^{42}$
T.~N.~Varga,$^{57,58}$
and J.~Weller$^{57,58}$
\begin{center} (DES Collaboration) \end{center}
}
}

\date{Accepted XXX. Received YYY; in original form ZZZ}

\pubyear{2021}

\begin{document}
\label{firstpage}
\pagerange{\pageref{firstpage}--\pageref{lastpage}}
\maketitle
%\linenumbers

\begin{abstract}
Recent cosmological analyses rely on the ability to accurately sample  from high-dimensional posterior distributions. A variety of algorithms have been applied in the field, but justification of the particular sampler choice and settings is often lacking. Here we investigate three such samplers to motivate and validate the algorithm and settings used for the Dark Energy Survey (DES) analyses of the first 3 years (Y3) of data from combined measurements of weak lensing and galaxy clustering. We employ the full DES Year 1 likelihood alongside a much faster approximate likelihood, which enables us to assess the outcomes from each sampler choice and demonstrate the robustness of our full results. 
We find that the ellipsoidal nested sampling algorithm \mnest{} reports inconsistent estimates of the Bayesian evidence and somewhat narrower parameter credible intervals than the sliced nested sampling implemented in \pchord{}.  
We compare the findings from \mnest{} and \pchord{} with parameter inference from the Metropolis--Hastings algorithm, finding good agreement.
We determine that \pchord{} provides a good balance of speed and robustness, and recommend different settings for testing purposes and final chains for analyses with DES Y3 data.  Our methodology can readily be reproduced to obtain suitable sampler settings for future surveys.
\end{abstract}


%%%%%%%%%%%%%%%%%%%%%%%%%%%%%%%%%%%%%%%%%%%%%%%%%%

%%%%%%%%%%%%%%%%% BODY OF PAPER %%%%%%%%%%%%%%%%%%

\section{Introduction}
The sampling of a posterior distribution is one of the central elements in current cosmological analyses. With the increasing complexity of cosmological surveys and the large amount of data available, it is a complicated challenge to extract cosmological parameters\footnote{
In this work, we use the term {\it parameters} to refer to the parameters characterizing a model, both nuisance and cosmological, for which we want to generate samples.  We use the term {\it hyperparameters} to refer to the parameters specific to sampler settings, which affect their performance, such as the number of samples we want to obtain, the stopping criteria, etc. This terminology is common in the machine learning literature. Note that the term {\it hyperparameters} can refer to different concepts, even in the field of cosmology  \citep{Lahav:1999hu, Hobson:2002zf, LuisBernal:2018drn}.
} 
because of the high dimensionality and complex shapes of the distributions. Nuisance parameters (accounting for various calibration and systematic effects) complicate the analysis by increasing the number of parameters well beyond the six of the standard \LCDM\ model of cosmology. 

Bayesian techniques give a principled framework for probabilistic inference, for instance characterising information about complex, usually non-Gaussian, posterior distributions for which the mean and standard deviation alone are insufficient to fully describe the shape of the distribution. Markov Chain Monte Carlo methods have traditionally been used for this purpose \citep{Metropolis:1953, Neal:1997}, and have a long history of applications in cosmology \citep[e.g.][]{Christensen:2001, Knox:2001, Lewis:2002, Verde:2003, Tegmark:2004, Dunkley:2005, Shaw:2007}. However, for some applications (such as model comparison and the comparison of different datasets) it is necessary to calculate not only the shape of the posterior distribution but also the Bayesian evidence. Nested Sampling \citep{Skilling:2006} is the method most commonly used for this purpose, because of its speed and its ability to obtain both the Bayesian evidence and the posterior distribution in the same calculation.

Because of their wide applicability, many tools have been developed to implement these sampling algorithms given a user-defined likelihood, and the choice to use one over another may be more driven by accessibility and ease of implementation than rigorous testing for the specific analysis at hand. 

As the constraining power of cosmological datasets has grown, 
different analyses have begun to diverge in their inferred parameter posteriors. Perhaps most famous is the discrepancy in the measurements of $H_0$  by the \Planck collaboration \citep{PlanckParameters:2018}  versus that obtained via distance ladder measurements \citep{riess2021comprehensive}, but there exists also tension between measurements of $S_8 \equiv \sigma_8(\Omega_{\rm m} / 0.3)^{0.5}$ from large-scale structure probes and that inferred by \Planck under \LCDM{} \citep[see ][for reviews on these tensions]{DiValentino:2020vvd, DiValentino:2020zio, Shah:2021onj}.
As these discrepancies could be indicators of new physics, it is vital that the inferences upon which they are based are robust to analysis choices such as the specific sampler and settings used. 

Most samplers include hyperparameters that allow one to tune the algorithm and, in the limit of infinite computing time and resources, allow one to obtain arbitrarily precise constraints. In practice, we require a balance of speed and accuracy, where it is feasible to run a large number of chains but the error introduced by the sampler is a negligible (or at least quantifiable) contribution to the analysis' error budget.
This is particularly true for the Dark Energy Survey \citep[DES,][]{DES:2005} combined weak lensing and galaxy clustering cosmology analysis (henceforth, 3x2pt), where the complexity of the data and analysis pipeline results in the need to run a large number of chains for validation purposes. 
In this work, we perform a careful investigation of several leading sampling algorithms available within the \cosmosis{} analysis framework, with a focus on \pchord{} and \mnest{} because of their ability to estimate the Bayesian evidence. We investigate how hyperparameters impact performance and focus particularly on avoiding biases in the parameter constraints and evidence, which could lead to mistaken interpretation of the core analysis results. We make recommendations for the sampler and settings for three different use cases of the DES Y3 data, which strike different balances of speed and accuracy.

There have been previous attempts at characterizing sampler performance. For example, \citet{Higson:2018cqj} developed diagnostic methods to assess errors from Nested Sampling chains, including the use of bootstrapping individual chain samples to assess uncertainty. We use some of these tools, but assess uncertainty using full independent chain realizations run over a wide range of parameter settings, and using high-resolution chains as benchmarks.
We combine tests on the first year of DES data (DES Y1) and on the results of a fast, approximate version of the likelihood that allows us to generate a large number  of sampling runs under the same hyperparameter settings. 

The paper is structured as follows: In \cref{sec:samplers} we introduce the methodology and notation of Bayesian parameter estimation, as well as the summary statistics that we will use throughout this work. In \cref{sec:method} we present the methodology and data used in this work. Our results are shown in \cref{sec:result}, and we present our conclusions in \cref{sec:conc}. All the data produced from this work is available upon request.

\section{Samplers}
\label{sec:samplers}

This section describes the formalism of parameter estimation in a Bayesian framework, as well as the three different sampling algorithms employed in this work. Detailed descriptions of the formalism can be found for example in \citet{Mackay:2002} and \citet{Sivia:2006}.

\subsection{The Bayesian framework}

In parameter estimation we have obtained some data $D$, we have assumed a theoretical model $M$, and we seek an estimate of the parameters $\theta$ of the model. This is accomplished by applying Bayes' theorem

\begin{linenomath}\begin{equation}
    P(\theta | D, M) = {P(D | \theta, M) \times P(\theta | M) \over P(D | M)}. 
\end{equation}\end{linenomath}
The quantities in this equation are usually labelled as
\begin{linenomath}\begin{equation}
    \P = {\L \times \Pi \over \Z}, 
\end{equation}\end{linenomath}
where $\P$ is the posterior, $\L$ the likelihood, $\Pi$ the prior, and $\Z$ the marginal likelihood or Bayesian evidence. The latter can be expressed as

\begin{linenomath}\begin{equation}
\label{Bayes}
\Z = \int \L \times \Pi \ \rmd \theta.
\end{equation}\end{linenomath}
This is typically a complicated and high-dimensional integral. Because $\Z$ acts as a normalising factor that does not depend on the parameters, it often plays no role for parameter estimation. There are, however, other applications where the Bayesian evidence is fundamental; one such case is Bayesian model comparison. Here we have two competing theoretical models $M_A$ and $M_B$ and we want to know which of these models is preferred given some measured data $D$. For this we calculate the ratio
\begin{linenomath}\begin{equation}
    {P(M_A | D) \over P(M_B | D)} = {P(D| M_A) \over P(D | M_B)} \times {P(M_A) \over P(M_B)},
\end{equation}\end{linenomath}
where the equality follows from Bayes' theorem. The second factor on the right-hand side is the ratio of the prior beliefs in the two models. If there is no prior reason to prefer one model over the other, then this term is unity and hence disappears. The first factor on the right-hand side is the ratio of the Bayesian evidences for the two models. Therefore, under the assumption of equal prior beliefs in the two models, we can find which model is preferred by the data by calculating the ratio

\begin{linenomath}\begin{equation}
    \label{r}
    R \equiv {\Z_A \over \Z_B}.
\end{equation}\end{linenomath}
This quantity is called the Bayes factor. Bayesian evidences are also used to quantify tension between different datasets \citep{Marshall:2006}.

In addition to the Bayesian evidence of \cref{Bayes}, we will compute two more summary statistics from our chains, which contain important information about our problem. The first one is the Kullback--Leibler divergence \citep{Kullback:1951}, given by
\begin{linenomath}\begin{equation}
    \mathcal{D}_{KL} = \int \mathcal{P} (\theta) \log {\mathcal{P}(\theta) \over \pi (\theta)} \ \rmd \theta.
\end{equation}\end{linenomath}
The Kullback-Leibler divergence measures the information gain when going from the prior to the posterior distribution, measured in natural bits, or {\it nats}. The Kullback--Leibler divergence can be used amongst other things to calculate the {\it information} between two data sets, which in turn can be used to calculate the {\it Suspiciousness} \citep{Handley:2019a}, and quantify the concordance between the data sets in a way that does not depend on prior volumes. 

Our last summary statistic is the Bayesian Model Dimensionality (henceforth BMD), which provides an estimate of how many Gaussian dimensions are constrained by our data \begin{linenomath}\begin{equation}
d = 2 \int \mathcal{P} (\theta) \left( \log {\mathcal{P}(\theta) \over \pi (\theta)} - \mathcal{D}_{KL} \right)^2 \rmd \theta.
\end{equation}\end{linenomath}
\citet{Handley:2019b} discuss the advantages of characterizing dimensionality via the BMD as opposed to other commonly-used measures like the Bayesian Model Complexity \citep{Spiegelhalter:2002}, such as not relying on a single point estimator. Another advantage of the BMD is that it can be computed directly from both nested sampling and Markov Chain Monte Carlo chains as 

\begin{linenomath}\begin{equation}
     \frac{\tilde{d}}{2} = \left\langle {\left(\log \mathcal{L} \right)}^2\right\rangle_\mathcal{P} - {\left\langle \log \mathcal{L} \right\rangle }^2_\mathcal{P},
\end{equation}\end{linenomath}
where $\langle \cdot \rangle_{\mathcal{P}}$ indicates an average over the posterior distribution.

\subsection{Metropolis--Hastings}
\label{sec:mh}

Markov Chain Monte Carlo (MCMC) is one of the most widely used methods for sampling probability distributions. It consists of using chains in which each element depends only on the previous one, known as Markov Chains, to obtain samples from the target distribution. 
The Metropolis--Hastings algorithm \citep{Hastings:1970} (denoted MH in the following) is a common MCMC method, widely used in various fields such as statistical mechanics or as here, Bayesian inference.
Here we use MH in order to generate samples from the posterior distribution of the cosmological and nuisance parameters. 
Below we describe the fundamental aspects of MCMC algorithms in general, and MH in particular, 
as well as details of its implementation within this work. 

Note that we include the MH sampler primarily as a benchmark against which we can compare the parameter estimation results of the nested samplers that are the main focus of this work. We have not made a significant effort to optimize the MH sampler's speed and performance, so a fair assessment of its computational cost compared to \pchord{} and \mnest{} is beyond the scope of this paper.

\subsubsection{The Metropolis--Hastings algorithm}

The goal of MCMC algorithms is to return samples from a distribution that converges towards a unique stationary distribution $\pi(\theta)$ (where $\theta$ are the cosmological and nuisance parameters) of the target distribution, in this case the posterior $\P(\theta)$. 
Given the transition matrix $p_{ij}$ of a Markov chain, which corresponds to the probability of moving from state $i$ at time $t$ to state $j$ at time $t+1$, we thus have
\begin{linenomath}\begin{equation}
    \pi_j = \sum_j p_{ij}\pi_i.
    \label{eq:mh}
\end{equation}\end{linenomath}
We now need to construct such a transition matrix.

The MH algorithm proposed in \cite{Hastings:1970} does so by requiring  $p_{ij}$ and $\pi$ to satisfy the so-called detailed balance
\begin{linenomath}\begin{equation}
    \pi_{i} p_{ij} = \pi_{j} p_{ji}.
\end{equation}\end{linenomath}
In MH, the transition matrix is then defined as
\begin{linenomath}\begin{equation}
    %\frac{p_i}{p_j} = \sum_j p_{ij}\pi_i
    p_{ij} = q_{ij} \alpha_{ij}
\end{equation}\end{linenomath}
where $q_{ij}$ is the proposal distribution (corresponding to a proposed `jump' in parameter space) and $\alpha_{ij}$ the acceptance distribution (corresponding to accepting this `jump' or not), defined as
\begin{linenomath}\begin{equation}
    \alpha_{ij} = \min \left(1, \frac{p_j q_{ji}}{p_i q_{ij}}\right). 
\end{equation}\end{linenomath}

If a chain of samples is selected using this algorithm for a large number of steps, the density of their resulting distribution will follow the target distribution,

\textit{i.e.} the posterior $\P(\theta)$.

Depending on the initial point in parameter space and the provided proposal, some samples drawn at the beginning of this process should be discarded as they are not representative of the posterior distribution.
This period is called burn-in, where the accepted points may be far from the peak of the posterior, and can be minimized if starting at a point in parameter space closer to the best-fit value (see \cite{cosmomcmc} for a discussion on choices to limit the burn-in period).
It can be explored by plotting the posterior or parameter values as a function of step number (or over-plotting these values from chains that started at different points), where the burn-in corresponds to samples before these values converge around the typical set.

One potential way of speeding up MH algorithms often used in cosmology is to take advantage of the fact that some parameters, known as `fast', do not affect the slowest parts of the likelihood calculation, which in the case of cosmology often involve the transfer function or line-of-sight integration. These parameters can be decorrelated and sampled separately, making sampling nearly as fast as it would for the `slow' parameters alone \citep{Lewis:2013}. When fast and slow parameters cannot be fully decorrelated in principle, they can be sampled using `dragging' ~\citep{2005math......2099N}, which consists of `dragging' the fast parameters while keeping the slow ones fixed, leading to fast likelihood evaluations. Both of these methods are implemented in the Cobaya package ~\citep{Torrado:2020dgo}.

One of the difficulties of using MCMC algorithms such as MH is the lack of definitive criteria ensuring the chain has converged towards the target distribution. 
Several criteria for testing the convergence have been proposed \citep{An98stephenbrooks, sandip2003}. It is also useful to study the autocorrelation of the MH chains to verify that the samples are independent on scales much smaller than the chain length.
In the following, we will mainly use the Gelman--Rubin diagnostic which is derived from the method proposed in \cite{GelmanRubin:1992} to monitor the convergence of MH chains.
This diagnostic works by comparing parameter estimates from a number of independent chains. Specifically, adopting the standard notation, it estimates the potential scale reduction factor $\hat{R}$ for a given parameter $\theta$, defined as
\begin{linenomath}\begin{equation}
    \hat{R} = \frac{\hat{V}}{W},
\end{equation}\end{linenomath}
where $\hat{V}$ is the estimator of the variance of the parameter and $W$ is the average of the variance of $\theta$ within a chain (in the above expression, the impact of degrees of freedom defined in \cite{GelmanRubin:1992} is neglected).
$\hat{R} \simeq 1$ implies that the distribution of the sampled parameter is close to stationary; while this does not \textit{guarantee} that the chain has converged, it is a good indication of convergence for unimodal posteriors when $\hat{R}$ is nearing 1 for all parameters. A typical convergence criterion is to stop when $\hat{R}-1<0.1$.
When considering $M$ independent chains, the variance estimator $\hat{V}$ is defined as
\begin{linenomath}\begin{equation}
    \hat{V} = \frac{N-1}{N} W + \frac{M+1}{MN} B
\end{equation}\end{linenomath}
where  $N$ is the length of the chains 
and $B/N$ is an estimate of the variance of the parameter mean $\bar{\theta}$ across chains
\textit{i.e.}
\begin{linenomath}\begin{equation}
    B/N = \frac{1}{M-1} \sum_{i=1}^{M} (\bar{\theta}_i - \bar{\theta})^2.
\end{equation}\end{linenomath}

\subsubsection{Implementation}

For this study we use a simple version of the MH sampler implemented within \cosmosis{}. 
In the configuration used here the MH sampler uses a fast--slow scheme in which each parameter subspace uses a separate multivariate Gaussian proposal, with a $1/3$ chance of each proposed jump length being drawn instead from an exponential distribution, to better explore parameter tails. We oversampled the fast subspace by a factor of 5 and have 9 fast parameters. 
In typical use of this sampler, the proposal is initially set to an estimate of parameter covariances, then tuned at the start of the chain. During tuning, the estimated parameter covariances are replaced with those computed from the points sampled in the chain up to that point. 

For this particular study we set the initial proposal using a parameter covariance extracted from a finished high-quality \pchord\ chain, and because that was expected to be close to the target distribution, we did not tune the proposal. This choice was motivated primarily by simplicity, in that it allowed us to use the MH sampler without adjusting its hyperparameters. Variations of this setting could have been used, namely using a more approximate initial proposal estimate --- for example, using only the diagonal part of the parameter covariance --- and then tuning the proposal. These choices would be expected to produce the same posterior estimate, just over a longer period of time.

Note that using the MH sampler requires sampling a scaled version of the primordial power spectrum amplitude, $10^{9}A_S$. This is because the relative size of the unscaled $A_S$ values compared to other parameters is small, which causes the proposal covariance matrix to be ill-conditioned.

\begin{figure*}
	\includegraphics[width=1.75\columnwidth]{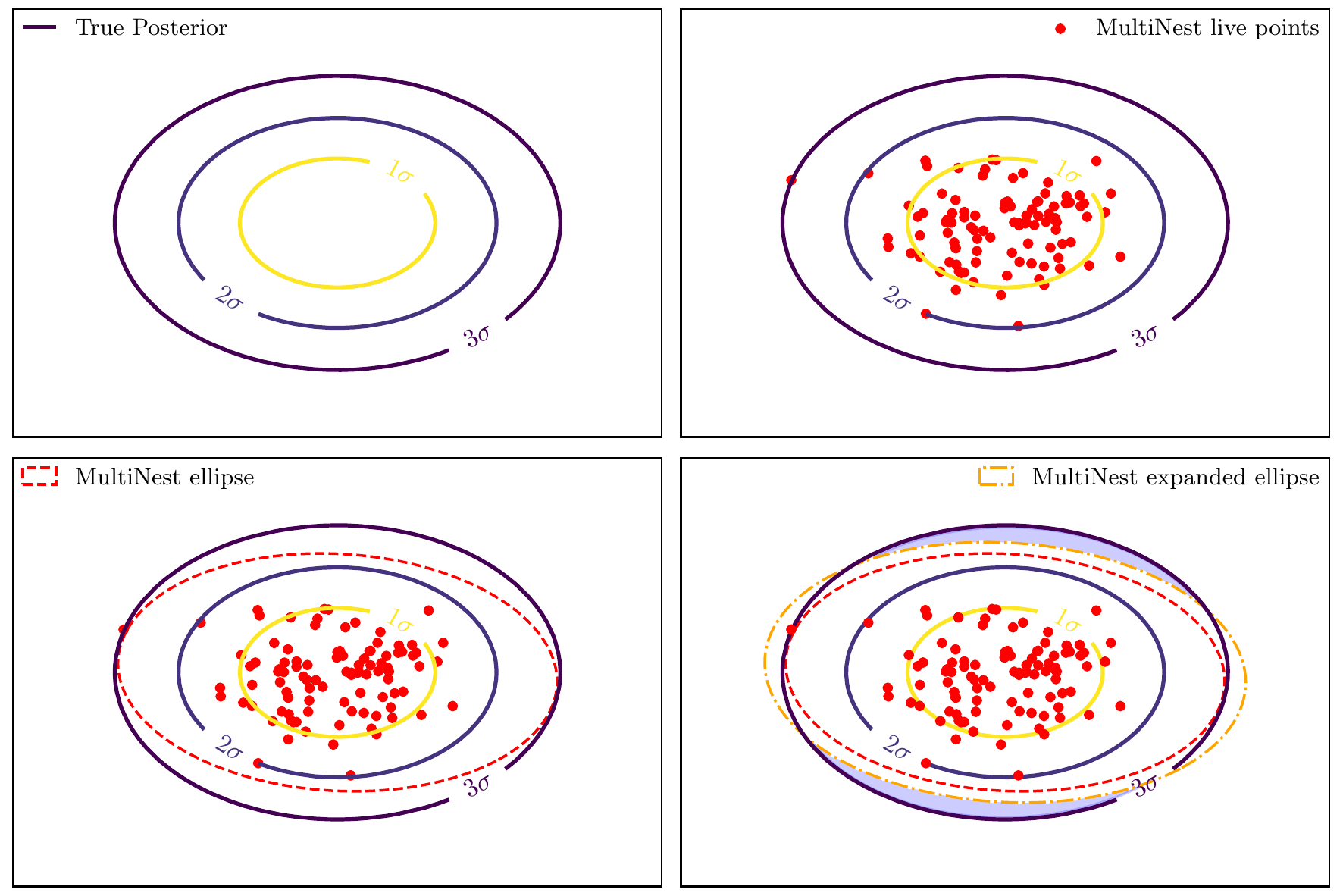}
    \caption{An example of \mnest's ellipsoidal sampling, and how it can lead to biases. When trying to sample a certain distribution (top left), \mnest\ randomly generates some points (top right). It then uses the covariance matrix obtained from those points point to calculate an ellipsoid enclosing all existing live points (bottom left, dashed line). That ellipsoid is expanded in volume by a factor inversely proportional to the efficiency, and samples are drawn from that ellipsoid (bottom right, dot-dashed line). As the latter plot shows in the light blue regions, if the magnification factor is not big enough (i.e.\ the efficiency is too high), this can lead to a bias in the estimation of the evidence.
    }
    \label{fig:multinest_example}
\end{figure*}

\subsection{MultiNest}

Multinest is an example of a nested sampling algorithm \citep{Skilling:2006} which, in contrast to MCMC samplers like MH, can be used to calculate the Bayesian evidence in addition to estimating the posterior.
Instead of selecting individual samples sequentially, nested sampling
starts with a large number of points (called `live' points), and then repeatedly selects the live point with the smallest value of the posterior density, eliminates it (turning it into a `dead point'), and then finds a new replacement live point with a posterior value larger than that of the point that was eliminated. The collection of all points (live and dead) can then be used to calculate the evidence while also serving as a (weighted) sample of the posterior. The most difficult part of Nested Sampling is finding new live points. It is extremely inefficient simply to randomly generate points until one with a higher posterior value is found (especially when most live points are close to the maximum of the posterior and when the problem has high dimensionality). This is the challenge that specific algorithms like \mnest{} are designed to address. 

\mnest\footnote{\url{https://github.com/farhanferoz/MultiNest}} is a publicly available code for Nested Sampling \citep{Feroz:2008, Feroz:2013}. It has been extensively used for cosmology analyses, including that of the first year (Y1) of  DES data \citep{DES-3x2:2018}. \mnest\ uses a technique called ellipsoidal sampling \citep{Mukherjee:2006}, where it calculates a $D$-dimensional ellipsoid from current set of live points, and finds the next point within that ellipsoid, expanded by a certain factor. 
\mnest\ also includes a clustering algorithm to identify multiple peaks in the posterior distribution, allowing it to sample multimodal posteriors. This was its main improvement over the ellipsoidal nested sampling code {\tt CosmoNest} \citep{Mukherjee:2006, Mukherjee:2005tr, Parkinson:2006ku, Pahud:2006kv}. There are other examples of ellipsoidal Nested Sampling algorithms, such as {\tt nestle}\footnote{\url{http://kylebarbary.com/nestle/index.html}} and {\tt dynesty}\footnote{\url{https://dynesty.readthedocs.io/en/latest/}}~\citep{2020MNRAS.493.3132S}, which uses dynamic sampling whilst still relying on ellipsoidal nested sampling.

As described in~\cite{Skilling:2006}, the standard Nested Sampling approach calculates the evidences using the accepted samples, and using an approximation for the distribution of sampling weights. In addition to this calculation, \mnest\ produces an alternative calculation of the Bayesian evidence using \textit{Importance Nested Sampling} (henceforth INS). INS, first introduced in the context of Nested Sampling in \cite{Cameron:2013}, uses all likelihood evaluations to estimate the evidence, instead of using only the accepted points in the \mnest\ run (which in some cases has acceptance rates as low as $\sim 1 \%$). In an ideal case, both estimates of the evidence should agree. In this work, when we refer to the \mnest{} evidence, we are referencing the `default' evidence calculation, and we will explicitly make reference to the INS evidence when that is not the case.

While ellipsoidal nested sampling leads to 
fast sampling, it can also lead to biases in both the posterior and the evidence estimation, as discussed later in the paper. 
This is illustrated by \cref{fig:multinest_example}: If the ellipsoid is not expanded enough, the calculation of the evidence will `miss' parts of the distribution.
These issues are discussed using \mnest\ as an example, but apply to any implementation of ellipsoidal nested sampling.

\subsection{PolyChord}
\label{sec:samplers_pchord}

An alternative code for Nested Sampling is \pchord\ \citep{Handley:2015a, Handley:2015b}.\footnote{\url{ https://github.com/PolyChord/PolyChordLite}} The difference between this algorithm and \mnest\ is in the approach to generating new live points. Instead of the ellipsoidal sampling, it uses so-called slice sampling \citep{Aitken:2013}, 
where new live points are generated by taking a random slice through the parameter space that includes the current live point, and randomly generating new points until one with higher likelihood is found. The process is then repeated with the new point and a slice in a new random direction, for a user-defined number of repetitions (\nrepeats{}) until the candidate live point is sufficiently uncorrelated with the initial live point. 
In practice, the sample covariance of existing points is used to decorrelate and whiten the parameter space, such that slices are performed on an affine transformation of parameter space where the relevant likelihood width is  $\mathcal{O}(1)$ in each direction. This both simplifies and accelerates the generation of new samples. 

Like \mnest, \pchord\ has a clustering algorithm which allows it to sample multimodal posterior distributions. In addition, \pchord\ is compatible with the fast--slow parameter implementation used by the code CosmoMC \citep{Lewis:2002, Lewis:2013}, which provides a significant increase in speed for cosmological likelihoods. While it is slower than \mnest\ in obtaining posterior distributions and Bayesian evidences for the models studied here, we will show that it more reliably gives unbiased results. 

\subsection{Other samplers}

In this work, we focus on three sampling algorithms commonly used in cosmology (Metropolis--Hastings, ellipsoidal nested sampling and slice nested sampling). 
One common sampler that we do not implement is \texttt{Emcee}, which is an affine-invariant MCMC sampler that uses an ensemble of walkers to traverse the posterior and update the proposal distribution before applying standard Metropolis--Hastings acceptance criteria \citep{emcee2013}. We found that in the large-dimensional parameter spaces tested here, the samples generated by \texttt{Emcee} had high enough levels of correlation so as to require intractable runtimes. 
Coupled with the inability to apply convergence criteria like the Gelman--Rubin statistic to correlated walkers and the large amount of samples that need to be discarded as burn-in, we decided not to include it in this study.

There exist other algorithms that, while perhaps not yet as widely used in cosmology, could become more common in the future. 
{\tt Zeus}\footnote{\url{https://zeus-mcmc.readthedocs.io/en/latest/}}~\citep{zeus} is an implementation of ensemble slice sampling~\citep{ess} for MCMC, and has the advantage of not requiring tuning of any hyperparameters, thus providing a promising alternative to traditional MCMC algorithms. Other algorithms such as Hamiltonian Monte Carlo~\citep{2017arXiv170102434B} or the No-U-Turn Sampler algorithm~\citep{2011arXiv1111.4246H} have existed for some time, but require accurate derivatives, which cannot be accessed easily in current cosmological theory codes such as {\tt CAMB}.

\section{Methodology}
\label{sec:method}

The goal of this paper is to compare the performance of the previously introduced methods for cosmological analysis. In cosmology, we usually perform inference with about six to eight cosmological parameters, and a number of nuisance parameters used to model systematic uncertainties. The nuisance parameters are usually marginalised over for cosmological constraints, though they may also be interesting in their own right (e.g.\ constraining galaxy bias or the amplitude of intrinsic alignment of galaxies). Here we use the pipeline for the DES Year 1 3x2pt analysis, which has 20 nuisance parameters. We assume a $w$CDM cosmological model, which 
allows for a varying equation of state for dark energy.
We therefore constrain 7 cosmological parameters: $\left\{ \Omega_{\rm m}, \Omega_{\rm b}, h, n_{\rm s}, A_{\rm s}, w, \Omega_{\nu} h^2 \right\}$, giving a total of 27 parameters to be sampled. 

In practice, we use two different pipelines in our analysis. We use the public DES Y1 3x2pt likelihood implemented in the cosmological parameter estimation code \cosmosis\ \citep{Zuntz:2015}, which includes all the samplers used in this work. 
In addition 
we also use a \textit{fast likelihood} that employs several approximations to reduce the evaluation time by a factor of $\sim50$. Both of these pipelines are described below.

\subsection{Fast Likelihood Analysis}

The sampling methods described in this work can be slow, and in some cases we can only understand the effects of tuning different hyperparameters by repeating the sampling a large number of times. For that purpose, we generated a {\it fast likelihood}, which produces posterior distributions that are similar to those of the DES Y1 pipeline, but uses multiple approximations to significantly reduce the run time. The resulting likelihood is an approximation to the true likelihood that allows for a large number of chains to be run and thus for the variance of samplers to be characterized. It can be considered a toy model that is substantially more applicable to our use case than the analytic models (e.g. Gaussian mixture models) that are often employed to characterize sampler behaviours.

The primary changes in the fast likelihood are:
\begin{enumerate}
    \item Using the fitting function presented in \cite{1998ApJ...496..605E} for the transfer function when computing the linear matter power spectrum;
    \item Acceleration of the calculation of the \texttt{Halofit} nonlinear scale (Equation A4 in \citealt{2012ApJ...761..152T}) using a non-iterative interpolation-based root-finding algorithm and trapezoidal integration;
    \item Calculation of the lensing efficiencies and Limber  angular correlation functions (Equations IV.3-IV.6 in \citealt{DES-3x2:2018}) using a simplified trapezoidal integration scheme.
\end{enumerate}

\subsection{Application to DES Y1 data}

We apply all the samplers described above to the DES 3x2pt analysis, running  \mnest{} and \pchord{} with a large number of different hyperparameter settings. 
Because the bulk of the work presented in this paper was performed while the analysis pipeline for the recently-released Y3 analysis~\citep{DES:2021wwk} was being developed, these tests are run using the DES Y1 data \citep{DES-3x2:2018} and the Y1 version of the DES modelling pipeline.
These data consist of a combination of three two-point correlation function measurements: Cosmic shear, galaxy--galaxy lensing, and galaxy clustering.

There are mainly two purposes to this paper: to find sampler settings that yield unbiased results for the DES analysis whilst minimizing the running time, and to generally understand the causes of bias in the parameter estimation or evidence calculation. 
The results presented in this work depend heavily on the dimensionality of the likelihood, as well as the form of the likelihood, and so cannot be generalized to all sampling problems. However, as most cosmological sampling problems have similar dimensionality and characteristics, these results should still be useful in guiding sampler choices in future cosmological analyses.

\section{Results}
\label{sec:result}

In this work, we have explored different sampling settings, to compare their performance and run time. Unless stated otherwise, all runs use the same likelihood, priors, and data, and are run using the same computing platform (the Cori system at NERSC) and with the same number of nodes. 

\subsection{Posterior validation with Metropolis--Hastings}
MCMC methods are expected to produce more reliable posteriors than Nested Sampling, because their convergence criterion is based on the posterior, not on the Bayesian evidence (which is difficult to estimate well from standard MCMC chains). Given this, before looking in detail at the effects of hyperparameters, we compare constraint contours from MH and \pchord{} in order to benchmark the accuracy of the nested sampling posterior estimates.

\begin{figure}
	\includegraphics[width=0.99 \columnwidth]{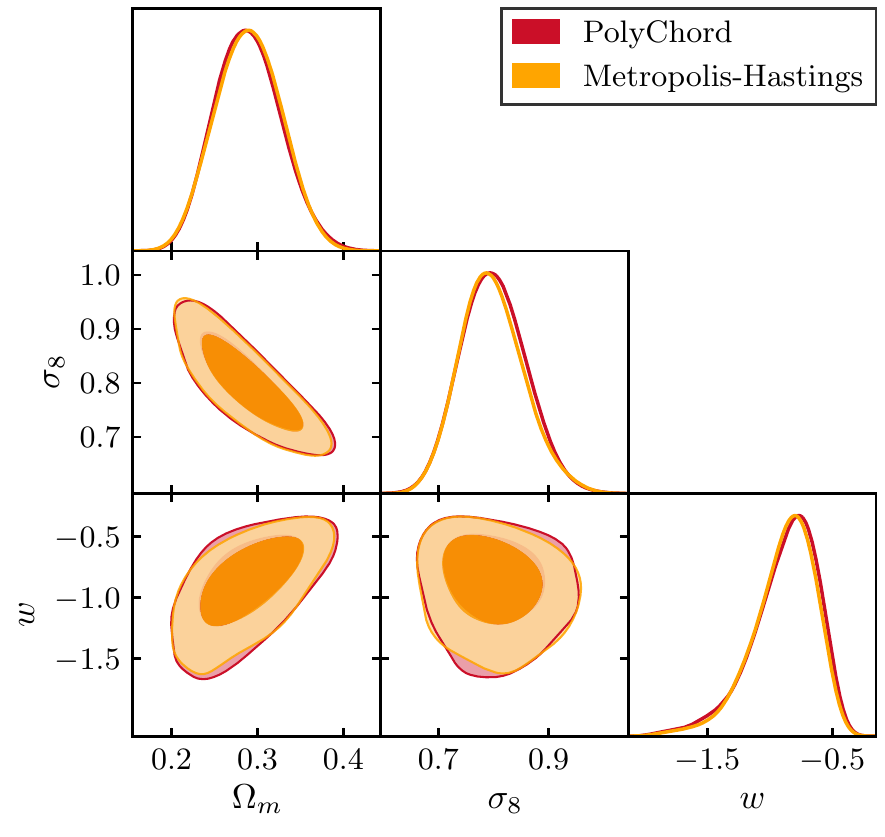}
    \caption{Posterior distribution for a high quality Nested Sampling run (red) and a Metropolis-Hastings run (yellow) which uses a full proposal covariance.
    }
    \label{fig:mh_compare}
\end{figure}

We run 8 MH chains in parallel using 4 processors per chains, spread across 2 nodes.
We stopped the chains once $\hat{R} - 1 < 0.02$ for all parameters (see Sec. \ref{sec:mh} for a description of the Gelman--Rubin statistic $\hat{R}$), amounting to 762,000 samples. 
We burn the first $20\%$ of the chains, as described in~\cref{sec:mh}.
Figure \ref{fig:mh_compare} shows the posterior estimated with these two sets of chains on the cosmological parameters $w$, $\Omega_{\rm m}$ and $\sigma_8$ along with the posterior estimated from a high quality \pchord\ chain.

We note that our MH run was slower than most nested sampling runs, using around 4,600 CPU-hours (6 days of walltime). Nested sampling is known to scale better with dimensionality, so this is on some level expected. However, we emphasize that   this is not necessarily a fair comparison because, as was discussed in \cref{sec:mh}, our MH runs did not employ a number of speed-up techniques which would likely be used in practice if MH were being used as the main sampler in an analysis. 
This is fine for our purposes, because as noted above we are using the MH chains to compare posterior distributions, not runtime. 

We interpret the good agreement observed in \cref{fig:mh_compare} between MH and the high quality \pchord\ chain as confirmation that with good enough settings, Nested sampling can accurately sample the posterior. We then explore what these settings need to be for both \mnest\ and \pchord.

\subsection{Multinest}
\label{sec:mnest}
\begin{table*}
	\centering
	\caption{Comparison of time required and output values for MultiNest with different settings. All runs use the DES Y1 3x2pt likelihood and a wCDM cosmology, with $128$ cores. The settings ${\tt OMP threads} = 1$ corresponds to 128 MPI threads, while ${\tt OMP threads} = 4$ corresponds to 32 MPI threads. CE refers to `Constant efficiency'. L Evals is the number of Likelihood evaluations. INS refers to the Importance Nested Sampling reported by \mnest, $D_{KL}$ is the Kullback--Leibler divergence, and BMD is the Bayesian Model Dimensionality. Reported uncertainties are via bootstrap resampling as computed by \texttt{Anesthetic} \citep{anesthetic19}.}
	\label{tab:multinest}
	\begin{tabular}{ccccc||rcc||cccc} % four columns, alignment for each
		\hline
		\nlive & Eff & Tol & {\tt OMP} & CE & Time (hrs) & Acceptance & L Evals & $\log Z$ & INS $\log Z$ & $D_{KL}$ & BMD\\
		\hline
		$675$ & $1$ & $0.3$ & $1$ & F & 11.2 & $0.067$ & $265314$ & $-277.71 \pm 0.17$&  $-285.14 \pm 0.05$ & $18.54 \pm 0.16$ & $13.0 \pm 0.3$\\
		$675$ & $1$ & $0.1$ & $1$ & F & 13.8 & $0.056$ & $340783$ & $-277.68 \pm 0.17$ & $-284.93 \pm 0.22$ & $18.61 \pm 0.15$ & $13.5 \pm 0.4$ \\
		$675$ & $1$ & $0.01$ & $1$ & F & 20.3 & $0.042$ & $510583$ & $-277.63 \pm 0.17$&  $-284.94 \pm 0.14$ & $18.42 \pm 0.15$ & $13.0 \pm 0.4$\\
		$675$ & $1$ & $0.1$ & $4$ & F & 46.3 & $0.053$ & $356248$ & $-277.53 \pm 0.17$&  $-285.09 \pm 0.06$ & $18.33 \pm 0.16$ & $12.8 \pm 0.4$\\
		$675$ & $1$ & $0.1$ & $1$ & T & 3.8 & $0.228$ & $78561$ & $-275.41 \pm 0.16$ & $-286.36 \pm 0.19$ & $17.59 \pm 0.15$ & $12.3 \pm 0.3$\\
		$250$ & $0.3$ & $0.1$ & $1$ & F & 4.3 & $0.053$ & $134554$ & $-278.68\pm 0.28$ & $-285.46 \pm 0.27$ & $19.32 \pm 0.28$ & $13.3 \pm 0.6$\\
		$675$ & $0.3$ & $0.3$ & $1$ & F & 14.6 & $0.054$ & $346242$ & $-278.52 \pm 0.17$ & $-285.01 \pm 0.14$ & $19.16 \pm 0.16$ & $13.5 \pm 0.4$\\
		$675$ & $0.3$ & $0.1$ & $1$ & F & 16.2 & $0.048$ & $396284$ & $-278.37 \pm 0.17$ & $-285.13 \pm 0.10$ & $19.14 \pm 0.17$ & $12.6 \pm 0.4$\\
		$675$ & $0.3$ & $0.01$ & $1$ & F & 21.4 & $0.040$ & $531284$ & $-278.53 \pm 0.17$ & $-284.63 \pm 0.28$ & $19.18 \pm 0.16$ & $13.7 \pm 0.4$\\
		$675$ & $0.3$ & $0.1$ & $4$ & F & 50.3 & $0.050$ & $383602$ & $-278.45 \pm 0.17$ & $-285.13 \pm 0.04$ & $19.21 \pm 0.16$ & $13.0 \pm 0.3$\\
		$675$ & $0.3$ & $0.1$ & $1$ & T & 4.6 & $0.199$ & $94746$ & $-276.56 \pm 0.17$ & $-285.78 \pm 0.24$ & $18.64 \pm 0.16$ & $12.3 \pm 0.4$\\
		$250$ & $0.1$ & $0.1$ & $1$ & F & 5.9 & $0.041$ & $178614$ & $-279.09 \pm 0.28$ & $-285.69 \pm 0.04$ & $19.98 \pm 0.28$ & $13.4 \pm 0.7$\\
		$675$ & $0.1$ & $0.1$ & $1$ & F & 23.7 & $0.035$ & $562784$ & $-278.88 \pm 0.17$ & $-285.16 \pm 0.10$  & $19.44 \pm 0.17$ & $12.9 \pm 0.4$\\
		%$1250$ & $0.1$ & $0.1$ & $1$ & F & 1d-06:31:09 & $0.037903$ & $977969$ & $279.109 \pm 0.13$ & $-284.85 \pm 0. 0.15$  &  & \\
		$675$ & $0.1$ & $0.1$ & $1$ & T & 6.9 & $0.106$ & $189995$ & $-278.34 \pm 0.17$ & $-285.27 \pm 0.06$ & $19.93 \pm 0.16$ & $13.2 \pm 0.4$\\ 
		$250$ & $0.01$ & $0.1$ & $1$ & F & 11.7 & $0.026$ & $294202$ & $-280.78 \pm 0.29$ & $-285.67 \pm 0.02$ & $20.97 \pm 0.30$ & $14.4 \pm 0.7$\\
		$675$ & $0.01$ & $0.1$ & $1$ & F & 39.3 & $0.025$ & $825487$ & $-280.62 \pm 0.18$ & $-285.38 \pm 0.02$ & $21.04 \pm 0.17$ & $13.6 \pm 0.4$\\
		$675$ & $0.01$ & $0.1$ & $1$ & T & 28.1 & $0.027$ & $754351$ & $-280.76 \pm 0.18$ & $-285.23 \pm 0.02$ & $20.96 \pm 0.16$ & $14.0 \pm 0.4$\\
		%$1250$ & $0.01$ & $0.1$ & $1$ & T & 2d-03:41:29 & $0.025418$ & $1528719$ & $-280.70 \pm 0.13$ & $-285.285 \pm 0.019$ &  & \\
		$250$ & $0.001$ & $0.1$ & $1$ & F & $39.0$ & $0.010$ & $823090$ & $-281.78 \pm 0.30$ & $-285.93 \pm 0.01$ & $21.62 \pm 0.28$ & $14.6 \pm 0.7$\\
		$675$ & $0.001$ & $0.1$ & $1$ & F & 109.5 & $0.010$ & $2116032$ & $-282.01 \pm 0.18$ & $-285.29 \pm 0.02$ & $21.99 \pm 0.17$ & $14.9 \pm 0.4$\\
		$675$ & $0.001$ & $0.1$ & $1$ & T & 126.3 & $0.008$ & $2756262$ & $-280.92 \pm 0.19$ & $-285.52 \pm 0.09$ & $20.17 \pm 0.17$ & $15.2 \pm 0.4$\\
		\hline
	\end{tabular}
\end{table*}

\mnest\ has several hyperparameters that can be tuned. These changes can increase accuracy in computing different quantities, at the expense of computing time. 
\cref{tab:multinest} shows timing and summary statistic results\footnote{Reported uncertainties are via bootstrap resampling as computed by the \texttt{Anesthetic} software package \cite{anesthetic19}.} of running \mnest{} chains with the DES likelihood using a variety of different choices for the sampler's hyperparameters.  We briefly describe these \mnest{} hyperparameters, (for more details see \cite{Feroz:2008, Feroz:2013}, henceforth F08, F13 respectively), and how they affect the sampler performance.

\begin{itemize}
    \item \nlive: the number of live points. This quantifies how many points are used to sample the posterior and is proportional to the expected final number of samples in the chain. 
    For the full likelihood we compare two different values: ${\tt n_{live} = 250}$ and ${\tt n_{live} = 675}$. The latter value is based on the fiducial 
    choice of ${\tt n_{live}} = 25*D$ where $D$ is the number of parameters being sampled.
    
\begin{figure}
	\includegraphics[width=0.99 \columnwidth]{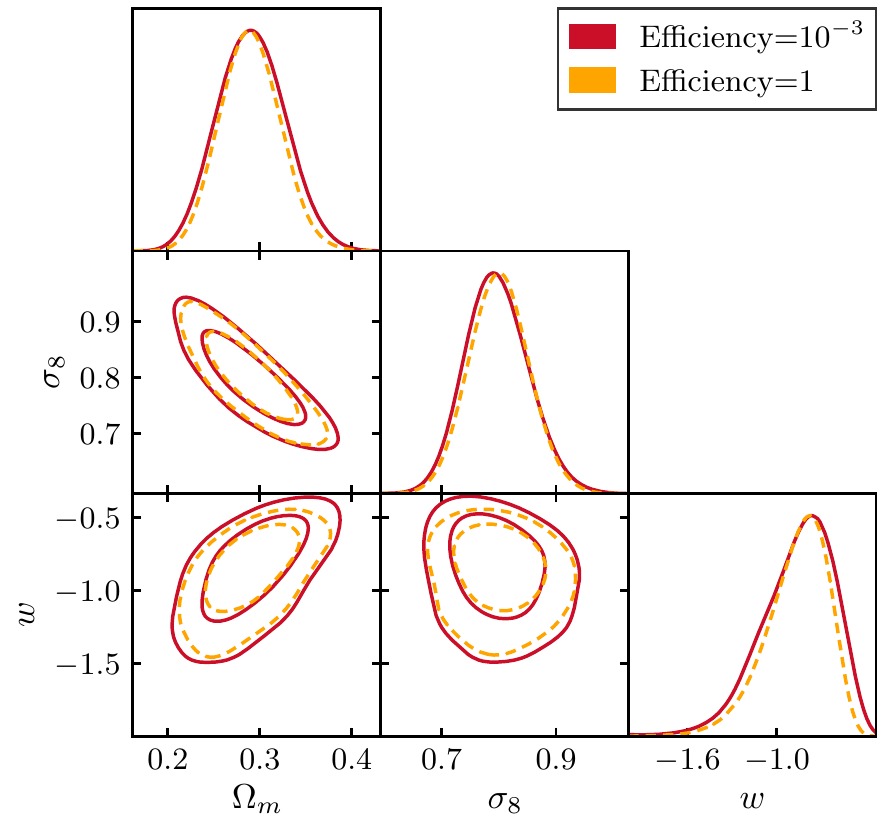}
    \caption{Marginalised one- and two-dimensional posterior distributions for two extreme values of the efficiency on \mnest. In red, a low value, which is therefore more likely to have fully sampled the tails of the distribution, and in the yellow dotted contours, a high value. We can see how the high efficiency does not fully sample the tails of the posterior distributions, and therefore gets narrower contours.}
    \label{fig:multinest}
\end{figure}

    A higher number of live points increases the accuracy 
    of the estimated posterior distributions and Bayesian evidence; we find that increasing \nlive\ decreases the uncertainty in $\log Z$ by a factor $\sqrt{\Delta \mathrm{n_{live}}}$, and increases run time linearly.

    \item \eff{}: a \mnest{}-specific hyperparameter that controls the size of the ellipsoids used by \mnest{} to search for new live points. To find the next live point after every step, \mnest\ uses the covariance of the existing live points to create an $N$-dimensional ellipsoid, then expands the ellipsoid by a factor of $1/{\tt efficiency}$ before using it to find the next live point.\footnote{The efficiency is thus a rough estimate of the acceptance rate, the probability that a point sampled from the expanded ellipsoid will have a higher likelihood than the point needing replacement. However the algorithm's acceptance scales better than this due to the ellipsoid hitting the prior boundaries in some of the parameter directions, as indicated by the fact that the BMD is less than the number of sampled parameters.} 
    This procedure is illustrated in \cref{fig:multinest_example}. 
 
 \begin{figure}
	\includegraphics[width=\columnwidth]{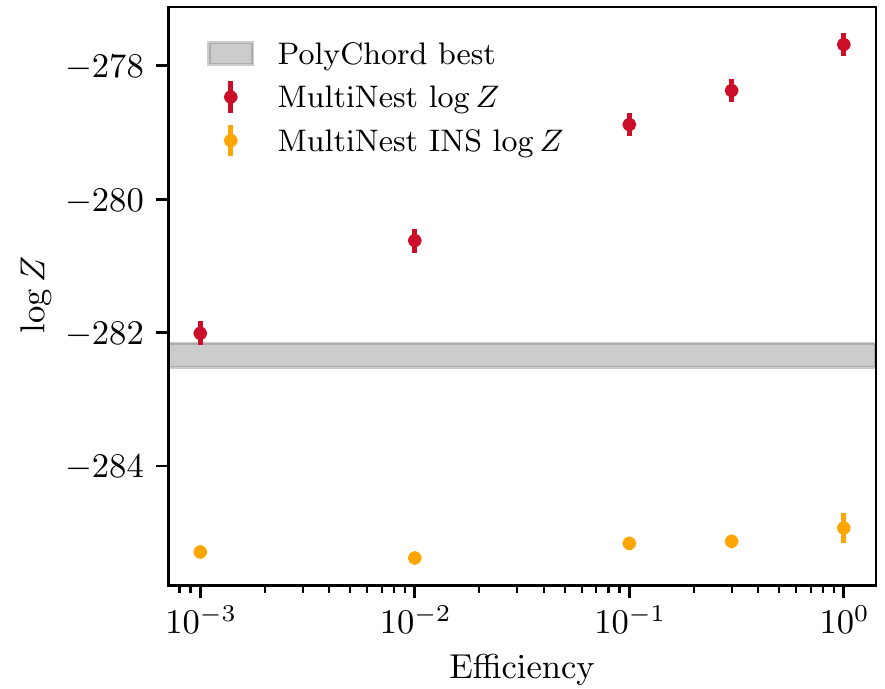}
    \caption{\mnest\ calculations of the evidence for different values of the efficiency. The \mnest\ values are plotted as red points, the \mnest\ INS evidence estimates in orange, and the grey band shows the 68\% confidence level of the best \pchord\ estimate.}
    \label{fig:efficiencies}
\end{figure}

    As previously explained, this Figure also shows a potential weakness of \mnest{}: we see that the expanded ellipsoid, shown with the dash-dotted orange line in the lower-right panel, is not capturing part of the tails of the true posterior distribution, shown in shaded blue. These regions will not be sampled, or considered when calculating the Bayesian evidence. This missing-posterior-tail bias will be more severe for higher values of efficiency, a finding that is reflected in the results shown in  \cref{tab:multinest}, \cref{fig:multinest}, and \cref{fig:efficiencies}.
    When the efficiency is too high, all the summary statistics calculated in this work are systematically wrong. Even for efficiencies as low as $10^{-3}$, we find a lack of convergence in summary statistics and disagreement with the best {\tt PolyChord} values. 
    
    \begin{figure*}
	\includegraphics[width=1.99\columnwidth]{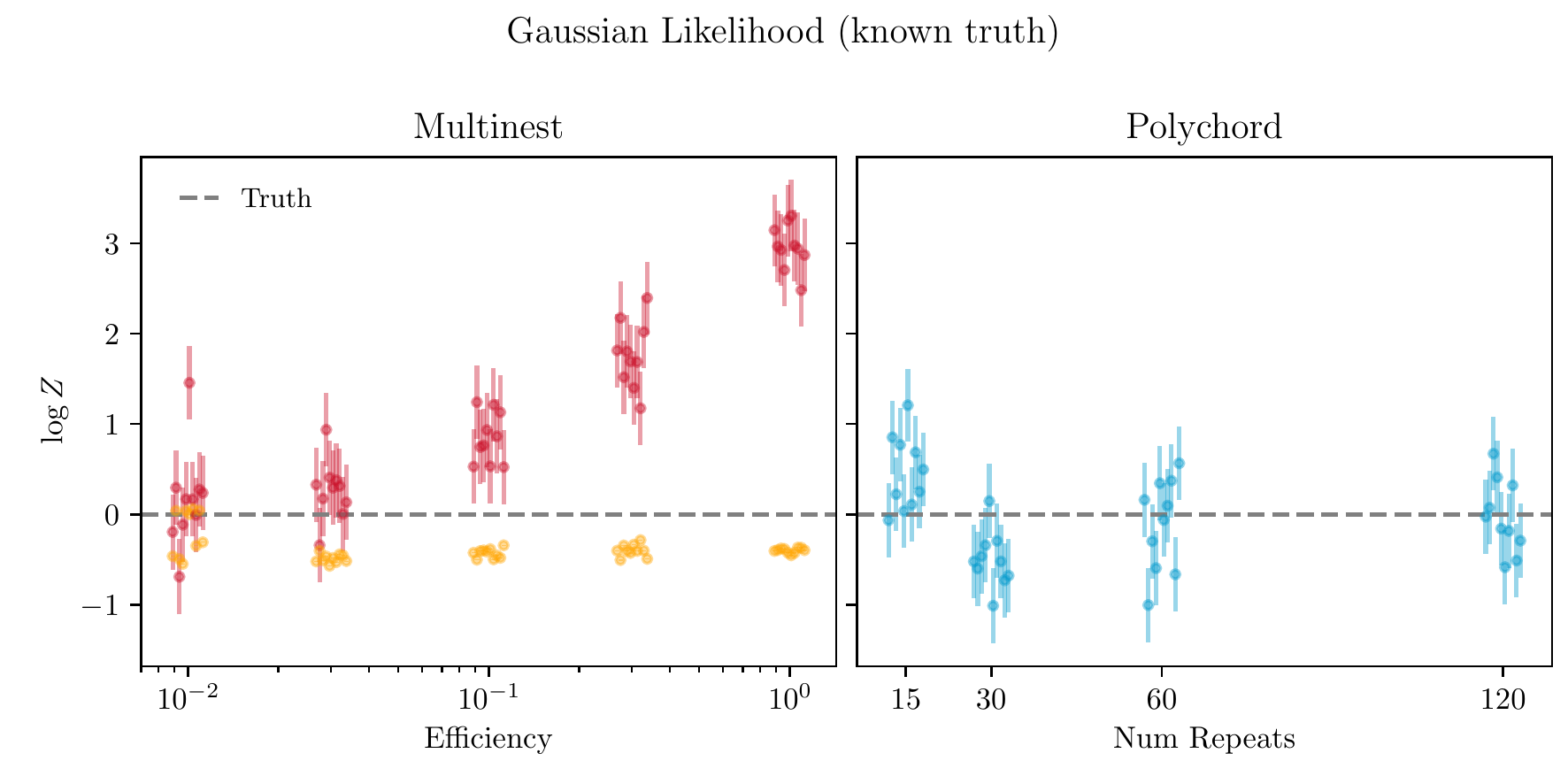}
    \caption{\mnest\ and \pchord\ estimations of the evidence for different values of the efficiency and \nrepeats\ parameters, for a Gaussian posterior distribution with known truth (dashed line). In both cases, we use ${\tt nlive} = 250$ and ${\tt tolerance} = 0.1$. For each setting, we show 10 different sampling runs, which we displace along the x axis for visualization purposes. The figure shows how \pchord\ gets more reliable evidence estimates even for low values of \nrepeats, where as \mnest\ gets biased estimates if the efficiency is not low enough. The yellow points include error bars, even though they are too small to be seen.
    }
    \label{fig:gaussian}
\end{figure*}

    This bias can be reproduced using a $27$-dimensional Gaussian posterior distribution, which has a known true evidence, as shown in~\cref{fig:gaussian}.
    Comparisons to this Gaussian toy model also show that \pchord\ gets more reliable evidence estimates, motivating us to adopt the best \pchord\ run on the DES likelihood as a benchmark `correct value'.

    We also see in \cref{tab:multinest} that there is an approximately power-law 
    relation between runtime and efficiency. Therefore, it can become extremely computationally expensive to achieve a low enough value of efficiency to obtain unbiased evidence estimates with \mnest{}.
    The importance of the efficiency hyperparameter for \mnest{} presents another challenge in that there is no principled way of knowing 
    what value of the efficiency should be used, or if the value used was low enough, without running the algorithm multiple times. 
    
    Note that biases from high efficiencies have a less severe impact on marginalized posteriors than on estimates of the Bayesian evidence. We can see this in \cref{fig:multinest}.
    While higher efficiency does cause the sampler to miss the tails of the distribution, even for very high values those missing tails are unlikely to significantly affect interpretation of the contours. Thus,  
    if we are only interested in the posterior distributions, we do not need to use efficiencies as low as would be needed if  we wanted to compute the Bayesian evidence. 
    
    \item {\bf Tolerance}: the stopping criterion. Both \mnest{} and \pchord{} can estimate how much the existing live points will contribute to the estimate of the evidence. When that contribution is smaller than the chosen value of the tolerance, the algorithm terminates. One can check whether the tolerance is low enough by plotting the progression of the weights of the chain, as shown in \cref{fig:tolerances}. 
     If the tolerance is low enough, this plot will show a peak that reaches unity, and will then decay back toward zero. A spike at the end shows that the contribution to the evidence from the final set of live points is too high, and the tolerance should be decreased.
    
    \cref{tab:multinest} shows that tolerance does not have a significant impact on either run-time nor on summary statistics. 
    Because of this, and because a chain initially run with higher tolerance can be resumed to reach a lower tolerance, the choice of this parameter is not considered a challenge: we simply recommend a tolerance that ensures that weights look similar to those on the right panel of \cref{fig:tolerances}.
    
    \item {\bf {\tt OMP} threads}: \mnest\ in \cosmosis\ uses a double parallelization scheme: The Boltzmann solver {\tt CAMB} \citep{Lewis:1999, Howlett:2012} is parallelized using {\tt OpenMP}, and the \mnest\ sampling algorithm uses {\tt MPI} parallelization. We tested two settings, both using the same number of nodes, but changing the number of cores used on each type of parallelization. We find that not using the {\tt OpenMP} parallelization greatly improves the sampling speed. We expected this, as \mnest\ will be faster when all cores are used by {\tt MPI} parallelization, up to the number of live points. As expected, changing this setting does not affect the results in any way apart from the run time. 
    
    \item {\bf Constant Efficiency}: \mnest\ can use a different sampling method, called `constant efficiency' mode. In this setting, we abandon the strategy of increasing the volume of the ellipse by a factor of $1/{\tt efficiency}$. Instead, the increase in the size of the ellipses changes at every step to match the input `constant efficiency' value in the sampling efficiency (i.e. the ratio of points accepted to points sampled). F13 describe how: 
    
    \smallskip
    
    {\it "Despite the increased chances of the fitted ellipsoids encroaching within the constrained likelihood volume (i.e. missing regions of parameter space for which $\mathcal{L} > \mathcal{L}_i$), past experience has shown (e.g. F08) this constant efficiency mode may nevertheless produce reasonably accurate posterior distributions for parameter estimation purposes."}
    
    \smallskip
    
    Our results agree with these statements in F13, with some caveats. Indeed, for efficiencies set to values of 0.3 and 0.1, constant efficiency mode produces significantly quicker runtimes and worse estimates of the evidence and other summary statistics than the standard mode. However, at lower efficiency values we find that this trend is inverted. For example, when the `constant efficiency' hyperparameter is set to $10^{-3}$, the constant efficiency runtime becomes longer than standard \mnest{}, and the evidence estimate also appears to converge to the correct value. However, given its longer runtime at efficiencies needed for accurate evidences, we do not recommend using `constant efficiency' to estimate the evidence. 
    
\end{itemize}

As previously discussed, \mnest\ produces an alternative Importance Nested Sampling evidence estimate. By examining \cref{tab:multinest} we can compare its dependence on hyperparameters to that of the main evidence calculation. 
This INS evidence estimate is very stable amongst all of our \mnest{} runs, %find very consistent results, 
always around the value of $\log Z \sim -285$, almost independently of the hyperparamter settings. 
\cref{fig:efficiencies} shows this graphically, with the INS estimates of the evidence in orange. 
They are significantly and consistently lower than the best estimate from \pchord. While this might suggest convergence to the `truth', this is belied by results from the Gaussian toy model of~\cref{fig:gaussian}, in which the \mnest{} INS evidence estimates are also systematically biased low. 
Our results thus appear to contradict the findings of F13, which  showed that in some toy models INS was more accurate than the baseline evidence estimate of \mnest.
 Note that in addition to being lower than the truth, the INS evidence reported also significantly underestimates its sampling error, i.e. the uncertainty caused by imperfect sampling.

\begin{figure*}
	\includegraphics[width=1.9 \columnwidth]{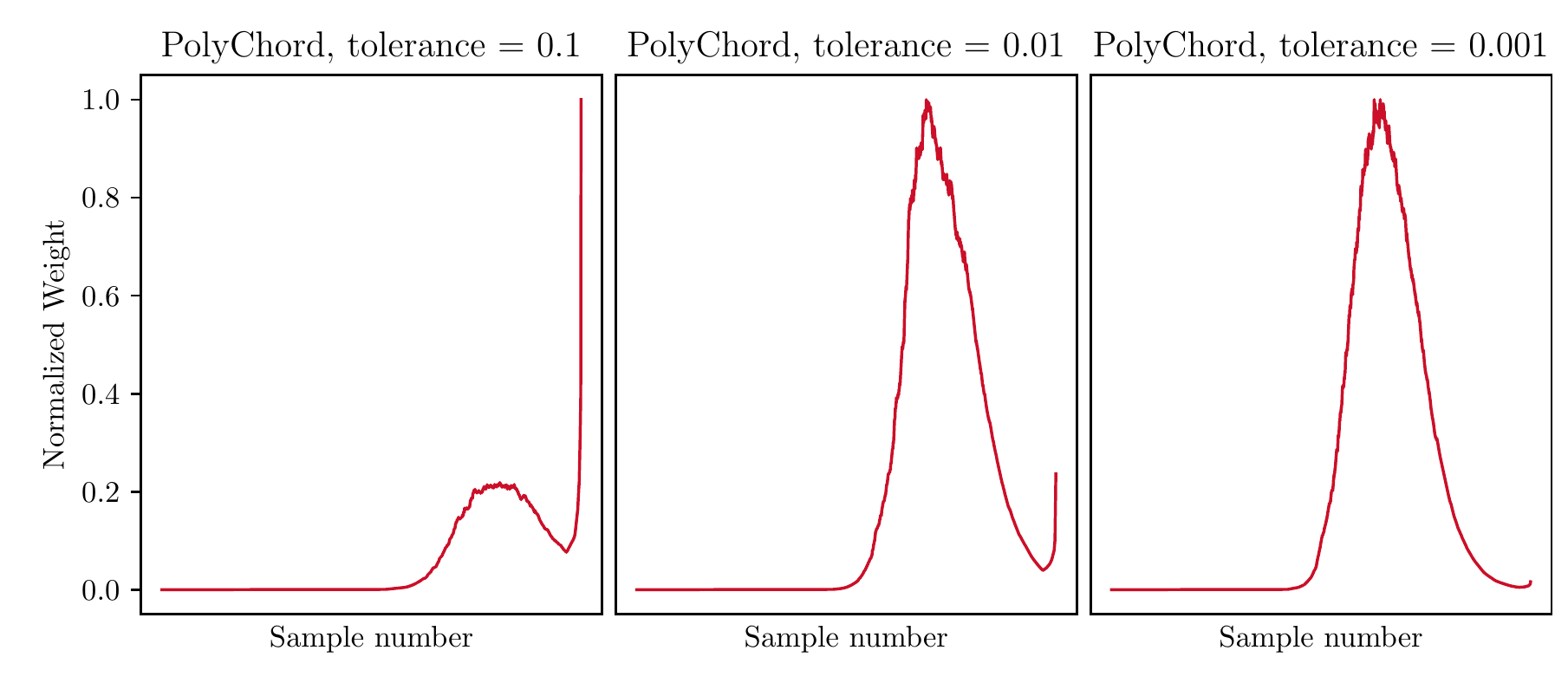}
    \caption{Normalized sample weight (weight divided by maximum weight) vs number of samples for different values of the tolerance. This plots serves as a convergence diagnostic; only the plot on the right has converged in this case. All plots use \pchord\ with \nlive = 250 and \nrepeats = 60.}
    \label{fig:tolerances}
\end{figure*}

\subsection{Polychord}
\label{sec:polychord}

\begin{table*}
	\centering
	\caption{Comparison of time required for \pchord\ with different settings. All runs use the DES Y1 3x2pt likelihood and a wCDM cosmology, with $128$ cores. The settings ${\tt OMP threads} = 1$ corresponds to 128 MPI threads, while ${\tt OMP threads} = 4$ corresponds to 32 MPI threads. } 
	\label{tab:polychord}
	\begin{tabular}{cccc||rcc||ccc} % four columns, alignment for each
		\hline
		\nlive & Tol & \nrepeats & {\tt OMP} & Time (hrs) & Acceptance & L Evals & $\log Z$ & $D_{KL}$ & BMD\\
		\hline
		$50$ & $0.1$ & $60$ & $1$ & 21.3 & $0.003$ & $473705$ & $-282.22 \pm 0.65$& $22.26 \pm 0.64$ & $12.1 \pm 1.3$\\
		$250$ & $0.3$ & $60$ & $1$ & 51.7 & $0.006$ & $1171509$ & $-282.01 \pm 0.29$ & $21.85 \pm 0.27$ & $14.3 \pm 0.7$\\
		$250$ & $0.1$ & $15$ & $1$ & 11.9 & $0.022$ & $342878$ & $-281.47 \pm 0.32$& $21.51 \pm 0.30$ & $14.3 \pm 0.7$\\
		$250$ & $0.1$ & $30$ & $1$ & 23.7 & $0.011$ & $675895$ & $-282.64 \pm 0.29$ & $22.35 \pm 0.27$ & $15.6 \pm 0.7$\\
		$250$ & $0.1$ & $60$ & $1$ & 46.2 & $0.006$ & $1319862$ & $-282.51 \pm 0.30$& $22.34 \pm 0.26$ & $15.2 \pm 0.7$\\
		$250$ & $0.1$ & $60$ & $4$ & 75.5 & $0.007$ & $1016058$ & $-282.48 \pm 0.29$& $22.26 \pm 0.29$ & $14.3 \pm 0.7$\\
		$250$ & $0.1$ & $120$ & $1$ & 87.4 & $0.003$ & $2597251$ & $-282.87 \pm 0.30$& $22.30 \pm 0.32$ & $14.9 \pm 0.8$\\
		$250$ & $0.03$ & $60$ & $1$ & 60.9 & $0.006$ & $1379582$ & $-282.46 \pm 0.30$& $22.21 \pm 0.30$ & $14.4 \pm 0.7$\\
		$250$ & $0.01$ & $60$ & $1$ & 62.4 & $0.006$ & $1504244$ & $-282.72 \pm 0.30$& $22.72 \pm 0.34$ & $13.5 \pm 0.6$ \\
		$250$ & $0.001$ & $60$ & $1$ & 68.3 & $0.005$ & $1670676$ & $-282.09 \pm 0.30$& $22.12 \pm 0.29$ & $14.2 \pm 0.6$\\
		$675$ & $0.1$ & $15$ & $1$ & 22.5 & $0.026$ & $733506$ & $-281.54 \pm 0.18$& $21.38 \pm 0.18$ & $14.1 \pm 0.4$\\
		$675$ & $0.1$ & $30$ & $1$ & 47.5 & $0.014$ & $1458873$ & $-282.52 \pm 0.19$& $22.32 \pm 0.19$ & $14.6 \pm 0.4$\\
		$675$ & $0.1$ & $60$ & $1$ & 117.9 & $0.007$ & $2863702$ & $-282.09 \pm 0.18$& $21.95 \pm 0.17$ & $14.4 \pm 0.5$\\
		$675$ & $0.01$ & $60$ & $1$ & 131.2 & $0.007$ & $3289309$ & $-282.14 \pm 0.18$& $21.79 \pm 0.17$ & $14.5 \pm 0.4$\\
		$675$ & $0.1$ & $120$ & $1$ & 191.4 & $0.003$ & $5795180$ & $-282.34 \pm 0.18$&  $22.32 \pm 0.17$ & $14.3 \pm 0.4$\\
			\hline
	\end{tabular}
\end{table*}

\begin{figure}
	\includegraphics[width=\columnwidth]{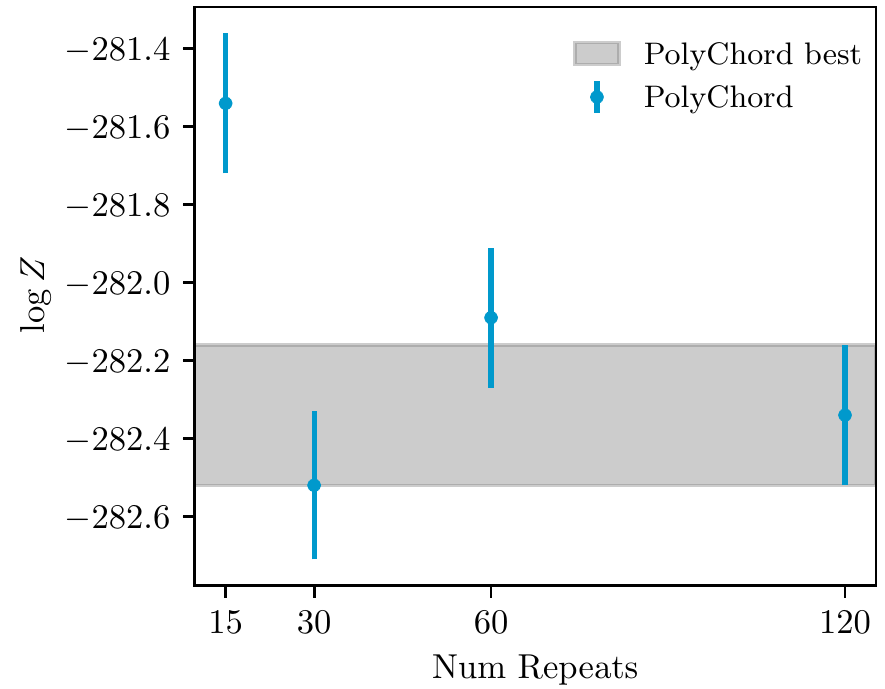}
    \caption{\pchord\ estimations of the evidence for different values of \nrepeats. The different values are plotted as red points, and the blue band shows the 68\% confidence level of the best \pchord\ estimate.}
    \label{fig:nrep}
\end{figure}

Like \mnest{}, \pchord{} has a number of hyperparameters which can be adjusted to balance running time and the accuracy of estimates for posterior distributions and summary statistics. The result of varying them is shown in~\cref{tab:polychord}.

\begin{itemize}
    \item \textbf{\nlive}, the number of live points. As with \mnest, more live points leads to an increase in the accuracy of the posterior distribution, and to a decrease in the error estimate for the evidence. Also, as was the case for \mnest, \pchord\ run-times scale linearly with the number of live points. 
    
    \item {\bf \tol{}}, the stopping criterion. It is defined in the same way for \cref{sec:mnest}, and the same conclusions apply: a lower tolerance does not have a significant impact on either runtime nor summary statistic accuracy. In fact, the tolerance seems to have even less of an impact on runtime for \pchord{} than for \mnest{}. 
    As before, we recommend using the weight-vs-step-number convergence diagnostic  illustrated in the right panel of \cref{fig:tolerances} to select a reasonable tolerance for a chain, and note that a \pchord{} chain can be resumed to reach a lower tolerance. 
    
    \item \nrepeats, the number of repeats. This hyperparameter is specific to \pchord's slice-sampling algorithm described in \cref{sec:samplers_pchord}. Recall that at every step, \pchord{} repeats the process of creating a slice through parameter space in a random direction, in which it finds a new potential live point. The value of \nrepeats\ dictates how many times this process is repeated for each sample selection. If this number is too low, the new live point will be correlated with the previous sample in the chain.
    Such correlations between live points could bias the posterior distribution and summary statistics. In this sense, a higher value of \nrepeats{} is somewhat akin to increasing the degree of `thinning' of a standard MCMC chain, with the result that more likelihood evaluations are performed but then discarded, in the interest of reducing systematic uncertainty. 
    
    The official \pchord\ paper~\citep{Handley:2015b} recommends using at least ${\tt n_{repeats}} \sim 2 D$, where $D$ is the number of dimensions being sampled. In this work, we tested values that are approximately $\left\{ 0.5, 1, 2, 4 \right\}$ times the number of dimensions. \cref{fig:nrep} shows the corresponding values of the summary statistics. We see that a value ${\tt n_{repeats}} = 15$ obtains biased estimates of the evidence for the full likelihood, but values as low as ${\tt n_{repeats}} = 30$ (for $D=27$) already obtain valid results. 
    The main conclusion is that, as expected, while a poor choice for \nrepeats\ can lead to biased results, the accuracy of estimates for evidence and other summary statistics are not nearly as sensitive to \nrepeats\ as they are to \mnest{}'s efficiency hyperparameter.
    
    \item {\bf {\tt OMP} threads}. As with  \mnest, we obtain the best results when we use all our cores for {\tt MPI} parallelization, up to the number of cores matching the number of live points.
\end{itemize}

\begin{figure*}
	\includegraphics[width=0.99 \columnwidth]{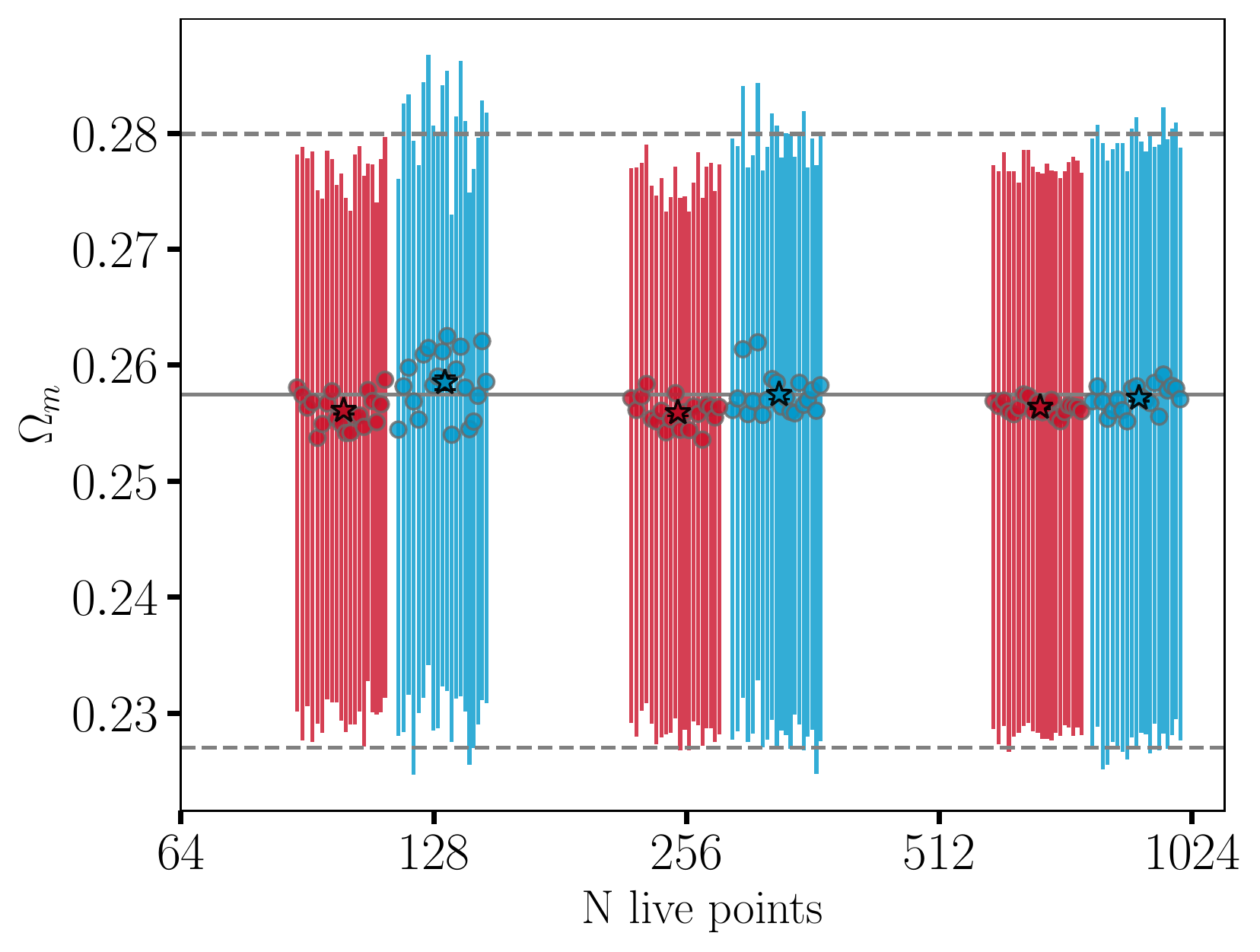}
	\includegraphics[width=0.99\columnwidth]{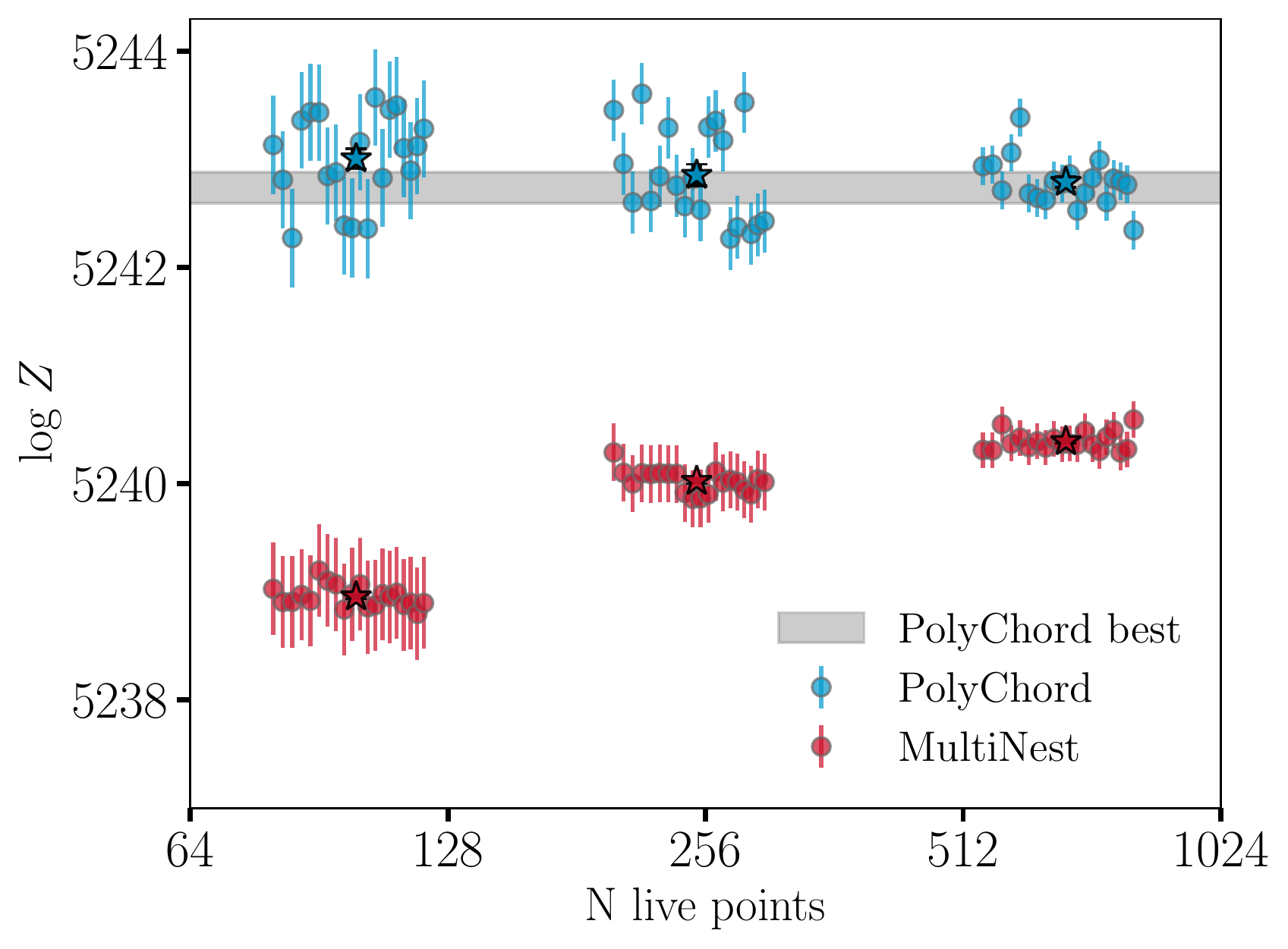}
% 	\columnwidth]{PC_MN_paramix0.pdf}
    \caption{\textit{Left}: 
     1D marginalized mean and $68\%$ credible intervals of $\Omega_{\rm m}$ for both \mnest{} (red)  and \pchord{} (blue) run using the fast likelihood at different values of nlive ($\{100, 250, 675\}$), while other hyperparameters are held constant at the default values listed in \cref{sec:FastLike}. 
     20 chains are run at each setting of \nlive{}, corresponding to the clusters of jittered points. The mean and standard error of chains at each setting are indicated by  stars at the center of each cluster.
     Horizontal lines correspond to the average mean and $1\sigma$ credible intervals of three high quality \pchord{} runs (\nlive{}=1000, \nrepeats{}=120, \tol=$10^{-3}$). There is good agreement between \pchord{} and  \mnest{} on the mean and small discrepancies between the credible intervals.  \mnest{} credible intervals are consistently smaller than those reported by \pchord{}. \cref{fig:s8_nlive_hist} shows this in greater detail for $S_8$.
    \textit{Right}:
  Estimates of the Bayesian evidence (and its sampler-reported uncertainty) for the same chains as the left-hand plot. The shaded band shows $\pm 1\sigma$ uncertainty on the mean evidence of the three high quality \pchord{} chains. The \pchord{} values are consistent at all settings, including at the lowest settings, with  each individual run consistent with the high-resolution ``truth" within its reported uncertainty. 
    In contrast, \mnest{} evidence estimates display a  systematic bias that is greater for small \nlive{}, and the reported uncertainty for individual chains is insufficient to make runs consistent across different values of \nlive{}. The reported uncertainties in $\log Z$ for each individual chain (given by the error bars) is consistent with the sampling variance across chain means for \pchord{}, but is greatly overestimated for \mnest{} where the means across chains are much more tightly clustered. This is shown more directly in  \cref{fig:fastnlivelogZhist}. 
    }
    \label{fig:fast_om_logz_nlive}
\end{figure*}

\subsection{Fast-likelihood tests of sampler variance}
\label{sec:FastLike}
While the studies above give us an indication of how changing the values of \pchord{} and \mnest{} hyperparameters affect runtime and the accuracy of summary statistic estimates, they do not tell us much about noise in those relations due to the particular realization of random points sampled in a given chain run. In order to assess this sampler variance, we run a large number of independent chains, which long runtimes make infeasible with the full DES likelihood studied above. Thus, to more robustly assess convergence properties and characterize how sampler variance changes across settings, we 
use the approximate fast likelihood to generate multiple chain realizations at each set of sampler hyperparameters.

We use a set of three `high-quality' \pchord{} chains with \nlive{} $=1000$, \nrepeats{} $=120$ and \texttt{tol} $=0.001$ to approximate the truth and compare with the performance of chains run with lower quality settings. Unless otherwise stated, we ran $20$ independent chains for each combination of settings.
While the previous sections illustrated the different impacts of the \eff{} and \nrepeats{} parameters (as the unique hyperparameters for each sampler), we found that varying \nlive{} had the most pronounced impact on posterior and evidence estimates for both \mnest{} and \pchord{}. This aligns with the recommendations of \cite{Higson:2018cqj}, who note that increasing \nlive{} is the most computationally efficient way to increase accuracy, as it decreases both stochastic and systematic contributions to the uncertainty. We therefore  mostly show results in this section with respect to varying \nlive{}. Unless otherwise stated, \mnest{} chains were run with \eff{} $=0.1$ and \tol{} $=0.1$, and \pchord{} chains with \nrepeats{} $=30$ and \tol $=0.01$.

\begin{figure}
	\includegraphics[width= \columnwidth]{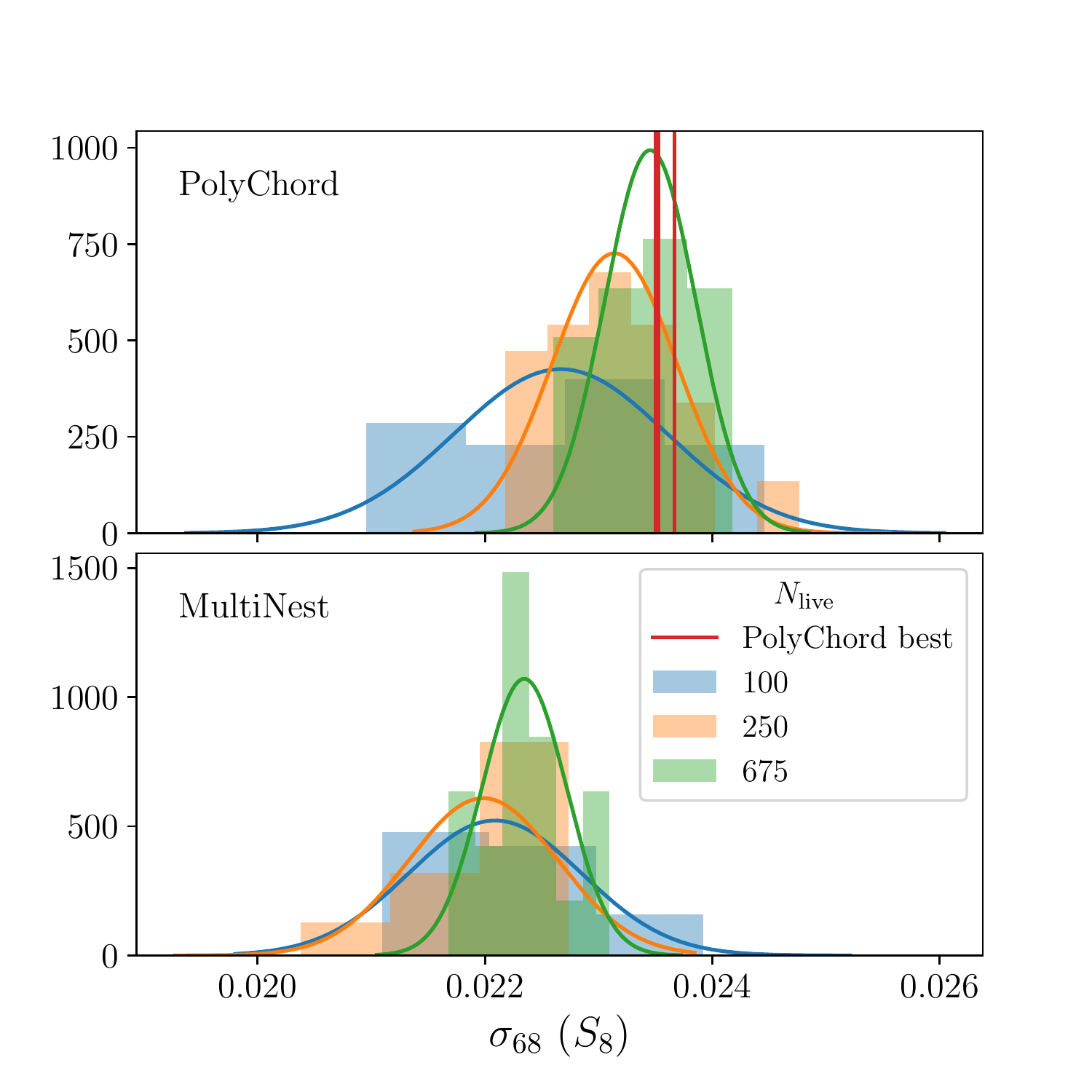}
    \caption{Histogram of the half-width of $68\%$ credible intervals for $S_8$ for many chains using the fast likelihood for both \pchord{} (top) and \mnest{} (bottom). \mnest{} systematically reports smaller credible intervals than \pchord{} for the range of settings tested, here shown with colors indicating different numbers of live points. Red lines indicate the credible intervals for the three high quality \pchord{} chains.}
    \label{fig:s8_nlive_hist}
\end{figure}

Fig.~\ref{fig:fast_om_logz_nlive} shows the marginalized 1D constraints on $\Omega_{\rm m}$ and log$Z$ for multiple independent \mnest{} (red) and \pchord{} (blue) chains with different numbers of live points. The average of the high-quality \pchord{} chains are shown in grey. Constraints on $\Omega_{\rm m}$ are consistent across the different of \nlive{} values for both samplers.
As expected based on results in previous sections, we find the $\Omega_{\rm m}$ credible intervals reported by \mnest{} to be consistently $\sim10\%$ smaller than those reported by \pchord{}. The same is true to a lesser extent for $S_8$. 
This is shown in Fig.~\ref{fig:s8_nlive_hist}, which depicts the estimated uncertainty in $S_8$  inferred from \pchord{} (top) and \mnest{} (bottom) chains run with different \nlive\ settings. The uncertainty is represented by the half-width of the $68\%$ credible interval ($\sigma_{68}(S_8)$), a quantity comparable to the standard deviation but more directly related to the marginalized quantities we are interested in for Bayesian cosmological inference and less sensitive to the tails of the posterior. As was noted in~\cref{sec:mnest}, underestimation of credible intervals is expected when the \mnest{} \eff{} parameter is too large. We see here that having low \nlive\ can also potentially cause parameter constraint error bars to be slightly underestimated.

Estimates of both the mean and dispersion of the Bayesian evidence differ significantly between the samplers. The values of $\log Z$ reported by \pchord{} are consistent across the range of settings we tested, indicating minimal systematic bias in the estimates due to hyperparameter settings. In contrast the reported \mnest{} evidence changes significantly as a function of \nlive, and to a degree much larger than the algorithm's reported uncertainty. This can also be seen in Figs.~\ref{fig:efficiencies} and \ref{fig:gaussian}, which show results for different likelihoods. 
As we increase \nlive{} in \cref{fig:fast_om_logz_nlive}, the \mnest{} evidence estimate shifts toward that reported by \pchord{}. This behavior suggests that the \mnest{} settings tested here are insufficient to obtain accurate evidences.

\begin{figure}
	\includegraphics[width= \columnwidth]{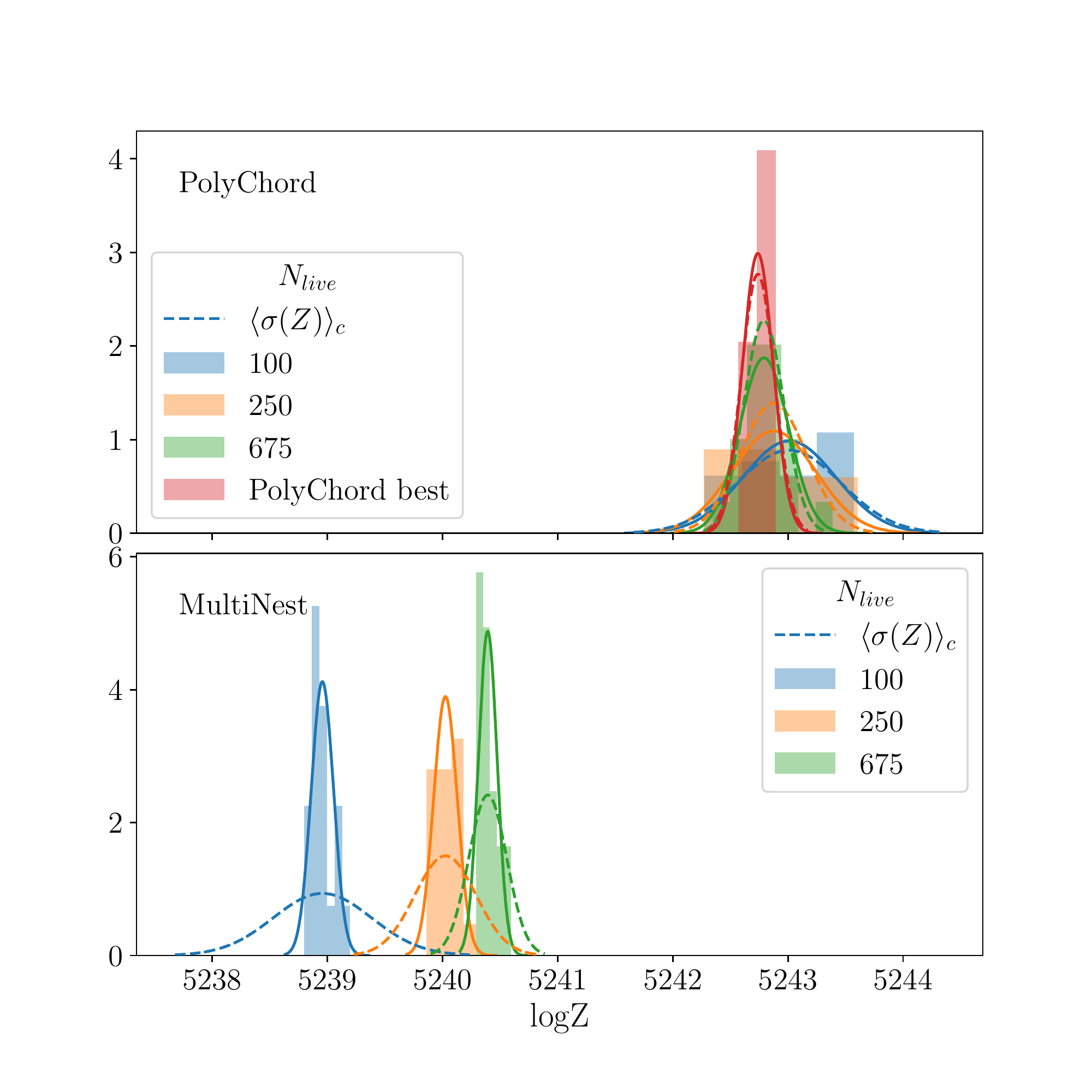}
    \caption{Histogram of reported $\log Z$ for many \pchord\ (top) and \mnest\ (bottom) chains run with different numbers of live points. Solid curves are Gaussian fits to the distribution in reported $\log Z$ across different chain realizations. Dashed Gaussian curves show the expected distribution based on the mean $\log Z$ and mean claimed uncertainty across chains. The relatively close agreement between solid and dashed curves indicates that the reported uncertainty in \pchord\ chains is fairly representative of the true sampling error. \mnest\ reports uncertainties considerably larger than the observed sampling variance. }
    \label{fig:fastnlivelogZhist}
\end{figure}

In addition to being robust to systematic shifts in the reported evidence, the sample variance in evidence estimates between different \pchord{} chains with the same settings shows good agreement with the sampler variance uncertainty reported from each individual chain. Fig.~\ref{fig:fastnlivelogZhist} shows the distribution of reported evidence values across the 20 chains at each value of \nlive{}. Dashed Gaussian curves are drawn according to
\begin{linenomath}\begin{equation}
    \mathcal{N}\left(\langle \log Z\rangle_{c}, \langle\sigma(\log Z)\rangle_c^2\right),
\end{equation}\end{linenomath}
where the averages are computed over each ensemble of 20 chains. For \pchord{}, these closely match the empirical distribution of reported evidences, indicated by the solid Gaussian curves which are fits to the histogram. 
In contrast, the reported statistical uncertainty in \mnest{} evidences is much greater than the observed  scatter across chain realizations. 

The uncertainty in $\log Z$ given by \pchord{} is more solidly grounded than that of \mnest{}, having been derived analytically by \citet{keeton2011}, whereas the \mnest{} evidence estimates use the relative entropy and are based on  information theoretic arguments. Our numerical results affirm the greater reliability of evidence errors from \pchord{}.

Despite \mnest{} overestimating the \textit{statistical} uncertainty in $\log Z$, the reported evidences are still inconsistent across hyperparameter settings, due to even greater \textit{systematic} shifts as \nlive{} is increased. However, none of the \mnest{} settings tested resulted in evidence values approaching the stable \pchord{} estimates.

% % Systematic shifts
We can also use this ensemble of chains to estimate and limit the contribution of sampler biases to the systematic error budget of the DES Y3 analysis.
The DES Y3 analysis requires that unmodeled systematics shift the maximum posterior point of key cosmological parameters by not more than $0.3\sigma$ in the 2D plane of $\Omega_{\rm m}$ and $S_8$ (c.f. \citealt{DES:2021rex}). 
Here we adopt a somewhat simplified but nonetheless strict criterion that the typical variation of parameter means across chain realizations is far below their statistical uncertainty. 
We thus require 
\begin{linenomath}\begin{equation}
    \sigma_c\left[\rm \bar{\theta}\right] < 0.1 \langle \sigma_s\left[\theta\right]\rangle_c,
\label{eq:sigmac_shift}
\end{equation}\end{linenomath} where the left-hand side is the standard deviation of the parameter mean across chain realizations and the right-hand side is a threshold, set equal to a fraction of the average of the  standard deviations of the parameter computed from the individual chains.  
Note that this requirement is closely related to requiring the Gelman--Rubin statistic defined in \cref{sec:mh} to be below a given threshold.

\begin{figure}
	\includegraphics[width=0.99 \columnwidth]{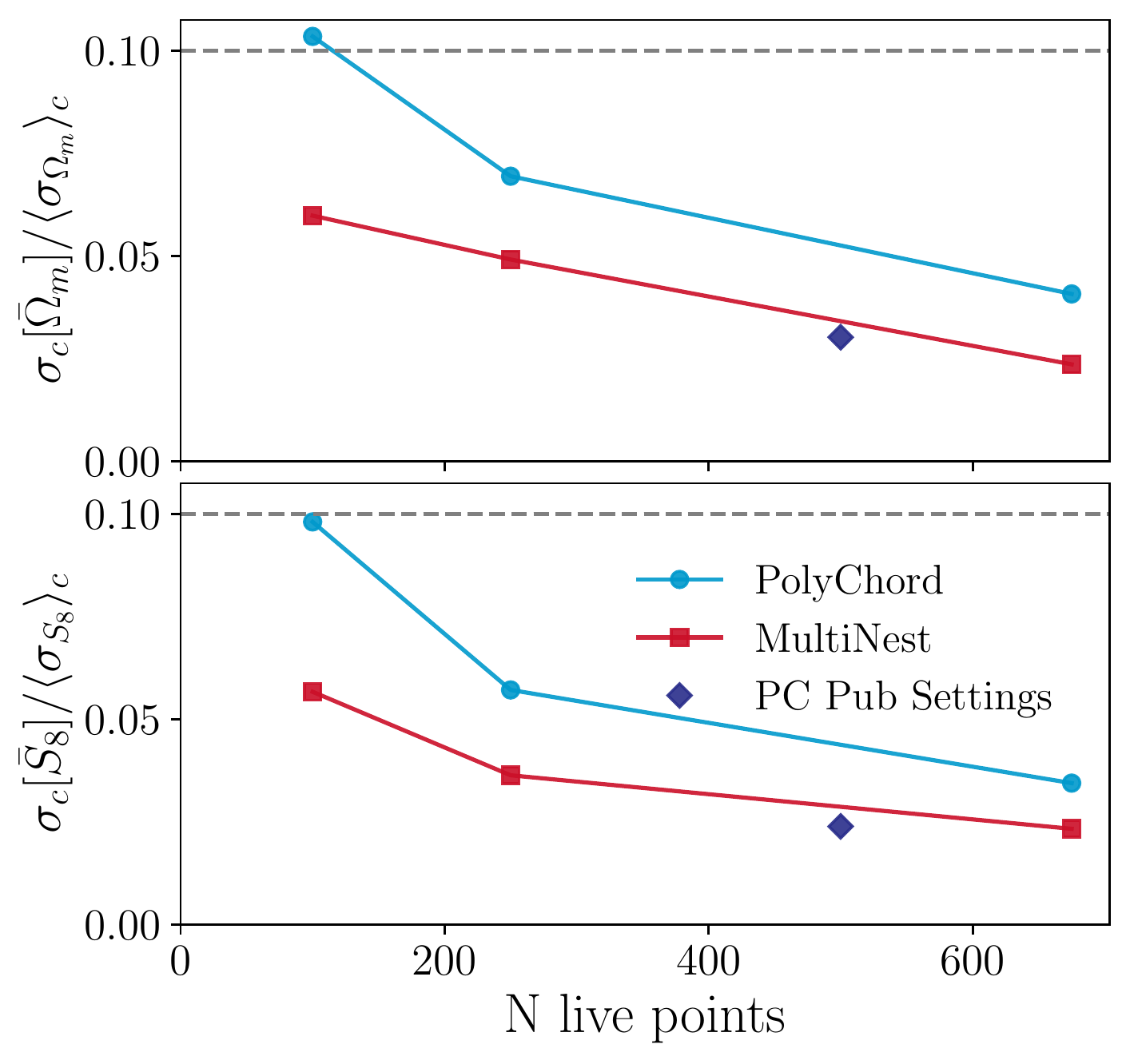}
    \caption{Standard deviation of parameter means across chains relative to the average within-chain parameter standard deviation. We require that the contribution to mean parameter shifts from sampler variance is small for settings used to run chains testing the impact of unmodeled systematics. The PolyChord publication settings can be found in~\cref{tab:recsettings}.}
    \label{fig:fastnliveR}
\end{figure}

We are primarily interested in shifts on $\Omega_{\rm m}$ and $S_8$, and so  require that any recommended sampler settings satisfy \cref{eq:sigmac_shift} for those parameters. \cref{fig:fastnliveR} confirms this requirement is fulfilled for \nlive$> 200$ with the fiducial values of the other hyperparameters. 

\section{Conclusions}
\label{sec:conc}

In this paper we have studied the performance of two commonly-used tools used to sample posteriors for cosmological analysis, \mnest{} and \pchord{}, as a function of their hyperparameter settings. Our analysis had two parts: testing multiple sampler settings on the DES Y1 3x2pt analysis to calibrate the time needed to get unbiased posterior distributions in the Y3 analysis, and also using a faster approximate version of the likelihood to characterize the amount of sampler variance and further validate those findings. 

We found that these Nested Sampling algorithms  require careful 
tuning of their hyperparameters, especially \nlive{} and \mnest{}'s ellipsoidal sampling efficiency. Particularly for \mnest{}, the wrong settings
can lead to a poor sampling of the tails of the posterior distribution, and to a biased evidence estimation. Furthermore, the superior speed of the  \mnest{} algorithm compared with \pchord{}'s slice sampling method is not present when sufficiently accurate sampling hyperparameters are used.
\pchord{} produces unbiased evidence estimates with reasonable settings, as well as contours that are in good agreement with those we find using Metropolis--Hastings. Therefore, our findings lead us to prefer \pchord{} over \mnest{}. 

The studies described in this paper were used to guide recommendations for sampler settings used for the DES Y3 cosmology analysis~\citep{DES:2021rex} as well as some Y1 follow-up papers~\citep{DES:2020mpv, DES:2020iqt}. These recommendations are summarized in Table~\ref{tab:recsettings} for three use-cases. We recommend that \mnest{} only be used for preliminary testing and pipeline debugging. While it is relatively fast, we found the (fairly standard) settings described in Case III of Table~\ref{tab:recsettings} to produce marginalized posterior widths for $\Omega_{\rm m}$ and $\sigma_8$ that are systematically underestimated by about $10\%$ and unreliable evidences. For most pre-publication testing, we recommend using the fast \pchord{} settings described in Case II. Those settings will produce unbiased posteriors and evidence values, though the resulting contours for marginalized posteriors will be somewhat noisy. Case I in Table~\ref{tab:recsettings} presents our recommendation for publication-quality results, with an increased number of live points extending the run-time but resulting in reduced noise in both posterior contours and evidence estimates.

While this study was performed specifically for DES Y3, our findings should be useful as a guide for cosmological analyses of similar dimensionality. Though our results were broadly consistent across three different likelihoods: the full DES Y1 likelihood, a fast approximate DES Y1 likelihood, and a 27D Gaussian toy model, the exact settings of~\cref{tab:recsettings} will likely need to be adjusted for problems where the number of dimensions or the shape of the posterior distribution change significantly. We found that good posterior and evidence estimates can be obtained with relatively low settings of \nrepeats{} $\sim D$ with \pchord{}, but that \mnest{} evidences were systematically biased except at extreme values of \eff{}, and \mnest{} missed tail regions of the posterior such that reported credible intervals were consistently $\sim10\%$ smaller than those reported by \pchord{}. We found that increasing \nlive{} had the greatest impact in improving accuracy of the summary statistics of interest.

Sampling algorithms are a key component of modern cosmological analyses and it is important to characterize their impacts on inference. As demonstrated in this work, poor choice of sampler and/or hyperparameter settings can lead to biased estimates of parameter constraints and other key summary statistics, but it is possible to achieve sufficiently unbiased estimates in realistic use cases. Through this work we motivate the sampling methods used for the DES Y3 analyses, and note that the fine margins demanded by precision cosmology will increasingly require heightened scrutiny of sampling tools.

\begin{table}
	\centering
	\caption{Recommended sampler settings based on this work, used for the DES Y3 cosmology analysis. Approximate wall-time estimates are given for a $w$CDM DES Y1 3$\times$2 chain run on 128 cores.}
	\label{tab:recsettings}
	\begin{tabular}{lc} 
		\hline\hline 
		\multicolumn{2}{c}{\bf Case I: Publication quality}\\
		\multicolumn{2}{c}{(one-time runs)}\\
		\hline
		Sampler & \pchord{}\\
		\nlive & 500 \\
		Tol & 0.01\\
		\nrepeats & 60 \\
		{\tt fast\_fraction} & 0.01\\
		Time to run:& $\sim4$ days\\
		\hline \hline
		\multicolumn{2}{c}{\bf Case II: Testing}\\
		\multicolumn{2}{c}{(noisy contours)}\\
		\hline
		Sampler & \pchord{}\\
		\nlive & 250 \\
		Tol & 0.1\\
		\nrepeats & 30 \\
		{\tt fast\_fraction} & 0.0\\
		Time to run:& $\sim1$ day\\
		\hline
		\hline
		\multicolumn{2}{c}{\bf Case III: Very preliminary results only}\\
		\multicolumn{2}{c}{(unreliable evidences, $\Omega_{\rm m}$ \& $\sigma_8$ posterior }
		\\\multicolumn{2}{c}{widths underestimated by $\sim 10$\%)}\\
		\hline
    	Sampler & \mnest{}\\
		\nlive & 250 \\
		Efficiency & 0.3 \\
		Tol & 0.1\\
		{\tt constant\_efficiency} & F\\
		Time to run:& $\sim6$ hours\\
		\hline\hline
	\end{tabular}
\end{table}

\section*{Acknowledgements}

% Individual acknowledgements
PL acknowledges STFC Consolidated Grants ST/R000476/1 and ST/T000473/1.

% software used

The analysis made use of the software tools {\sc SciPy}~\citep{Jones:2001}, {\sc NumPy}~\citep{Oliphant:2006},  {\sc Matplotlib}~\citep{Hunter:2007}, {\sc CAMB}~\citep{Lewis:1999,Howlett:2012}, {\sc GetDist}~\citep{Lewis:2019}, {\sc Multinest}~\citep{Feroz:2007,Feroz:2008,Feroz:2013},  {\sc Polychord}~\citep{Handley:2015a,Handley:2015b}, {\sc anesthetic}~\citep{anesthetic19}, and \cosmosis~\citep{Zuntz:2015}. Elements of the DES modeling pipeline additionally use {\sc Cosmolike}~\citep{Krause:2016jvl}, {\sc Halofit}~\citep{2012ApJ...761..152T,Bird:2011rb}, {\sc Fast-PT}~\citep{McEwen:2016fjn}, and {\sc Nicaea}~\citep{Kilbinger:2008gk}.

% computing clusters
This work was supported  through computational resources and
services provided by the National Energy Research Scientific Computing Center (NERSC), a U.S. Department of Energy Office of Science User Facility operated under Contract No. DE-AC02-05CH11231; and by the Sherlock cluster, supported by Stanford University and the Stanford Research Computing Center.

% standard DES acknowledgement
Funding for the DES Projects has been provided by the U.S. Department of Energy, the U.S. National Science Foundation, the Ministry of Science and Education of Spain, 
the Science and Technology Facilities Council of the United Kingdom, the Higher Education Funding Council for England, the National Center for Supercomputing 
Applications at the University of Illinois at Urbana-Champaign, the Kavli Institute of Cosmological Physics at the University of Chicago, 
the Center for Cosmology and Astro-Particle Physics at the Ohio State University,
the Mitchell Institute for Fundamental Physics and Astronomy at Texas A\&M University, Financiadora de Estudos e Projetos, 
Funda{\c c}{\~a}o Carlos Chagas Filho de Amparo {\`a} Pesquisa do Estado do Rio de Janeiro, Conselho Nacional de Desenvolvimento Cient{\'i}fico e Tecnol{\'o}gico and 
the Minist{\'e}rio da Ci{\^e}ncia, Tecnologia e Inova{\c c}{\~a}o, the Deutsche Forschungsgemeinschaft and the Collaborating Institutions in the Dark Energy Survey. 

The Collaborating Institutions are Argonne National Laboratory, the University of California at Santa Cruz, the University of Cambridge, Centro de Investigaciones Energ{\'e}ticas, 
Medioambientales y Tecnol{\'o}gicas-Madrid, the University of Chicago, University College London, the DES-Brazil Consortium, the University of Edinburgh, 
the Eidgen{\"o}ssische Technische Hochschule (ETH) Z{\"u}rich, 
Fermi National Accelerator Laboratory, the University of Illinois at Urbana-Champaign, the Institut de Ci{\`e}ncies de l'Espai (IEEC/CSIC), 
the Institut de F{\'i}sica d'Altes Energies, Lawrence Berkeley National Laboratory, the Ludwig-Maximilians Universit{\"a}t M{\"u}nchen and the associated Excellence Cluster Universe, 
the University of Michigan, NSF's NOIRLab, the University of Nottingham, The Ohio State University, the University of Pennsylvania, the University of Portsmouth, 
SLAC National Accelerator Laboratory, Stanford University, the University of Sussex, Texas A\&M University, and the OzDES Membership Consortium.

Based in part on observations at Cerro Tololo Inter-American Observatory at NSF's NOIRLab (NOIRLab Prop. ID 2012B-0001; PI: J. Frieman), which is managed by the Association of Universities for Research in Astronomy (AURA) under a cooperative agreement with the National Science Foundation.

The DES data management system is supported by the National Science Foundation under Grant Numbers AST-1138766 and AST-1536171.
The DES participants from Spanish institutions are partially supported by MICINN under grants ESP2017-89838, PGC2018-094773, PGC2018-102021, SEV-2016-0588, SEV-2016-0597, and MDM-2015-0509, some of which include ERDF funds from the European Union. IFAE is partially funded by the CERCA program of the Generalitat de Catalunya.
Research leading to these results has received funding from the European Research
Council under the European Union's Seventh Framework Program (FP7/2007-2013) including ERC grant agreements 240672, 291329, and 306478.
We  acknowledge support from the Brazilian Instituto Nacional de Ci\^encia
e Tecnologia (INCT) do e-Universo (CNPq grant 465376/2014-2).

This manuscript has been authored by Fermi Research Alliance, LLC under Contract No. DE-AC02-07CH11359 with the U.S. Department of Energy, Office of Science, Office of High Energy Physics.

%%%%%%%%%%%%%%%%%%%%%%%%%%%%%%%%%%%%%%%%%%%%%%%%%%



\bibliographystyle{mnras}
\bibliography{refs} % if your bibtex file is called example.bib


%%%%%%%%%%%%%%%%%%%%%%%%%%%%%%%%%%%%%%%%%%%%%%%%%%

%%%%%%%%%%%%%%%%% APPENDICES %%%%%%%%%%%%%%%%%%%%%

\section*{Affiliations}
$^{1}$ Department of Physics \& Astronomy, University College London, Gower Street, London, WC1E 6BT, UK\\
$^{2}$ Department of Physics and Astronomy, Pevensey Building, University of Sussex, Brighton, BN1 9QH, UK\\
$^{3}$ Department of Physics, University of Michigan, Ann Arbor, MI 48109, USA\\
$^{4}$ Lawrence Berkeley National Laboratory, 1 Cyclotron Road, Berkeley, CA 94720, USA\\
$^{5}$ Jodrell Bank Center for Astrophysics, School of Physics and Astronomy, University of Manchester, Oxford Road, Manchester, M13 9PL, UK\\
$^{6}$ Perimeter Institute for Theoretical Physics, 31 Caroline St. North, Waterloo, ON N2L 2Y5, Canada\\
$^{7}$ Jet Propulsion Laboratory, California Institute of Technology, 4800 Oak Grove Dr., Pasadena, CA 91109, USA\\
$^{8}$ Instituto de Astrof\'{\i}sica e Ci\^{e}ncias do Espa\c{c}o, Faculdade de Ci\^{e}ncias, Universidade de Lisboa, 1769-016 Lisboa, Portugal\\
$^{9}$ Department of Physics, Carnegie Mellon University, Pittsburgh, Pennsylvania 15312, USA\\
$^{10}$ Department of Physics and Astronomy, University of Pennsylvania, Philadelphia, PA 19104, USA\\
$^{11}$ Institute for Astronomy, University of Edinburgh, Edinburgh EH9 3HJ, UK\\
$^{12}$ Department of Astronomy, University of California, Berkeley,  501 Campbell Hall, Berkeley, CA 94720, USA\\
$^{13}$ Department of Astronomy/Steward Observatory, University of Arizona, 933 North Cherry Avenue, Tucson, AZ 85721-0065, USA\\
$^{14}$ Department of Astronomy, University of Geneva, ch. d'Ecogia 16, CH-1290 Versoix, Switzerland\\
$^{15}$ Laborat\'orio Interinstitucional de e-Astronomia - LIneA, Rua Gal. Jos\'e Cristino 77, Rio de Janeiro, RJ - 20921-400, Brazil\\
$^{16}$ Fermi National Accelerator Laboratory, P. O. Box 500, Batavia, IL 60510, USA\\
$^{17}$ CNRS, UMR 7095, Institut d'Astrophysique de Paris, F-75014, Paris, France\\
$^{18}$ Sorbonne Universit\'es, UPMC Univ Paris 06, UMR 7095, Institut d'Astrophysique de Paris, F-75014, Paris, France\\
$^{19}$ Faculty of Physics, Ludwig-Maximilians-Universit\"at, Scheinerstr. 1, 81679 Munich, Germany\\
$^{20}$ Kavli Institute for Particle Astrophysics \& Cosmology, P. O. Box 2450, Stanford University, Stanford, CA 94305, USA\\
$^{21}$ SLAC National Accelerator Laboratory, Menlo Park, CA 94025, USA\\
$^{22}$ Instituto de Astrofisica de Canarias, E-38205 La Laguna, Tenerife, Spain\\
$^{23}$ Universidad de La Laguna, Dpto. Astrof\'isica, E-38206 La Laguna, Tenerife, Spain\\
$^{24}$ Center for Astrophysical Surveys, National Center for Supercomputing Applications, 1205 West Clark St., Urbana, IL 61801, USA\\
$^{25}$ Department of Astronomy, University of Illinois at Urbana-Champaign, 1002 W. Green Street, Urbana, IL 61801, USA\\
$^{26}$ Institut de F\'{\i}sica d'Altes Energies (IFAE), The Barcelona Institute of Science and Technology, Campus UAB, 08193 Bellaterra (Barcelona) Spain\\
$^{27}$ Institut d'Estudis Espacials de Catalunya (IEEC), 08034 Barcelona, Spain\\
$^{28}$ Institute of Space Sciences (ICE, CSIC),  Campus UAB, Carrer de Can Magrans, s/n,  08193 Barcelona, Spain\\
$^{29}$ California Institute of Technology, 1200 East California Blvd, MC 249-17, Pasadena, CA 91125, USA\\
$^{30}$ Astronomy Unit, Department of Physics, University of Trieste, via Tiepolo 11, I-34131 Trieste, Italy\\
$^{31}$ INAF-Osservatorio Astronomico di Trieste, via G. B. Tiepolo 11, I-34143 Trieste, Italy\\
$^{32}$ Institute for Fundamental Physics of the Universe, Via Beirut 2, 34014 Trieste, Italy\\
$^{33}$ Observat\'orio Nacional, Rua Gal. Jos\'e Cristino 77, Rio de Janeiro, RJ - 20921-400, Brazil\\
$^{34}$ Hamburger Sternwarte, Universit\"{a}t Hamburg, Gojenbergsweg 112, 21029 Hamburg, Germany\\
$^{35}$ Santa Cruz Institute for Particle Physics, Santa Cruz, CA 95064, USA\\
$^{36}$ Institute of Theoretical Astrophysics, University of Oslo. P.O. Box 1029 Blindern, NO-0315 Oslo, Norway\\
$^{37}$ Kavli Institute for Cosmological Physics, University of Chicago, Chicago, IL 60637, USA\\
$^{38}$ Instituto de Fisica Teorica UAM/CSIC, Universidad Autonoma de Madrid, 28049 Madrid, Spain\\
$^{39}$ Department of Astronomy, University of Michigan, Ann Arbor, MI 48109, USA\\
$^{40}$ Excellence Cluster Origins, Boltzmannstr.\ 2, 85748 Garching, Germany\\
$^{41}$ School of Mathematics and Physics, University of Queensland,  Brisbane, QLD 4072, Australia\\
$^{42}$ Center for Cosmology and Astro-Particle Physics, The Ohio State University, Columbus, OH 43210, USA\\
$^{43}$ Department of Physics, The Ohio State University, Columbus, OH 43210, USA\\
$^{44}$ Center for Astrophysics $\vert$ Harvard \& Smithsonian, 60 Garden Street, Cambridge, MA 02138, USA\\
$^{45}$ Australian Astronomical Optics, Macquarie University, North Ryde, NSW 2113, Australia\\
$^{46}$ Lowell Observatory, 1400 Mars Hill Rd, Flagstaff, AZ 86001, USA\\
$^{47}$ Departamento de F\'isica Matem\'atica, Instituto de F\'isica, Universidade de S\~ao Paulo, CP 66318, S\~ao Paulo, SP, 05314-970, Brazil\\
$^{48}$ Department of Astrophysical Sciences, Princeton University, Peyton Hall, Princeton, NJ 08544, USA\\
$^{49}$ Instituci\'o Catalana de Recerca i Estudis Avan\c{c}ats, E-08010 Barcelona, Spain\\
$^{50}$ Physics Department, 2320 Chamberlin Hall, University of Wisconsin-Madison, 1150 University Avenue Madison, WI  53706-1390\\
$^{51}$ Institute of Astronomy, University of Cambridge, Madingley Road, Cambridge CB3 0HA, UK\\
$^{52}$ Centro de Investigaciones Energ\'eticas, Medioambientales y Tecnol\'ogicas (CIEMAT), Madrid, Spain\\
$^{53}$ School of Physics and Astronomy, University of Southampton,  Southampton, SO17 1BJ, UK\\
$^{54}$ Computer Science and Mathematics Division, Oak Ridge National Laboratory, Oak Ridge, TN 37831\\
$^{55}$ National Center for Supercomputing Applications, 1205 West Clark St., Urbana, IL 61801, USA\\
$^{56}$ Institute of Cosmology and Gravitation, University of Portsmouth, Portsmouth, PO1 3FX, UK\\
$^{57}$ Max Planck Institute for Extraterrestrial Physics, Giessenbachstrasse, 85748 Garching, Germany\\
$^{58}$ Universit\"ats-Sternwarte, Fakult\"at f\"ur Physik, Ludwig-Maximilians Universit\"at M\"unchen, Scheinerstr. 1, 81679 M\"unchen, Germany\\

%%%%%%%%%%%%%%%%%%%%%%%%%%%%%%%%%%%%%%%%%%%%%%%%%%

% Don't change these lines
\bsp	% typesetting comment
\label{lastpage}
\end{document}